# Intelligent fault diagnosis of worm gearbox based on adaptive CNN using amended gorilla troop optimization with quantum gate mutation strategy

Govind Vashishtha [1], Sumika Chauhan [1*], Surinder Kumar [2], Rajesh Kumar [2], Radoslaw Zimroz [1] and Anil Kumar [3]

[1] Faculty of Geoengineering, Mining and Geology, Wroclaw University of Science and Technology, Na Grobli 15, 50-421 Wroclaw, Poland

[2] Precision Metrology Laboratory, Department of Mechanical Engineering, Sant Longowal Institute of Engineering andTechnology, Longowal 148 106, India

[3] College of Mechanical and Electrical Engineering, Wenzhou University, Wenzhou 325 035, China

* Corresponding author, Email address: sumi.chauhan2@gmail.com

**Abstract:** The worm gearbox is a power transmission system that has various applications in industries. Being vital element of machinery, it becomes necessary to develop a robust fault diagnosis scheme for worm gearbox. Due to advancements in sensor technology, researchers from academia and industries prefer deep learning models for fault diagnosis. The optimal selection of hyperparameters (HPs) of deep learning models plays a significant role in stable performance. Existing methods mainly focused on manual tuning of these parameters, which is a troublesome process and sometimes leads to inaccurate results. Thus, exploring more sophisticated methods to optimize the HPs automatically is important. In this work, a novel optimization, i.e. amended gorilla troop optimization (AGTO), has been proposed to make the convolutional neural network (CNN) adaptive for extracting the features to identify the worm gearbox defects. Initially, the vibration and acoustic signals are converted into 2D images by the Morlet wavelet function. Then, the initial model of CNN is developed by setting hyperparameters. The search space of each HP is identified and optimized by the developed AGTO algorithm. The classification accuracy has been evaluated by AGTO-CNN, which is further validated by the confusion matrix. The performance of the developed model has also been compared with other models. It has been observed that the proposed AGTO not only achieved the highest degree of recognition accuracy i.e. 98.95 % but also achieved the least standard of deviation of 0.2145 than that of other classifiers. The AGTO algorithm is examined on twenty-three classical benchmark functions and the Wilcoxon test which demonstrates the effectiveness and dominance of the developed optimization algorithm. The results obtained suggested that the AGTO-CNN has the highest diagnostic accuracy, more stable while diagnosing the worm gearbox.

# 1. Introduction

One of the most challenging engineering tasks is comprehending fracture mechanics and recognizing failures. Determining the defective mechanical equipment and the cause of the defect is not an easy task as it involves a series of the tasks such as identification of the defect mechanism, monitoring of the faults, its early detection and prediction [1–3]. Condition monitoring is one of the techniques that can reduce the said troubles. In condition monitoring, the data is gathered and elucidated with the help of the most sophisticated sensors and data acquisition systems of recent times without stopping the machine. Based on these interpretations, it can be easily determined whether the machine component is defective or not so that the necessary action can be taken for accurate maintenance [4–7].

The rotating machinery is the key element of any industry. These types of machinery are powered through gears, bearings and other parts, which can become faulty during their operation and can affect the performance of the machines and even result in the breakdown of the machine. The gear system is the most common transmission element that is used in industries to transmit power most proficiently [8–10]. Based on the requirement of the power transmission and the working conditions, the gear system is classified into different categories, such as spur gear, bevel gear, helical gear and worm gear etc. The worm gear system gives a high speed reduction. The worm gear has a special mechanism which consists of a worm screw and worm wheel that meshes with each other. During harsh operating conditions, worm gears (both worm screw and worm wheel) are subjected to different types of failures, such as wear, surface fatigue and tooth breakage [11–13]. This necessitates the need for an appropriate fault diagnosis scheme to plan a suitable maintenance task based on the information gathered from data collected for the proper functioning of the gearbox.

Recently, researchers have widely used techniques based on artificial intelligence and machine learning due to the increment in the availability of data. Vashishtha and Kumar [14] proposed a robust and superior version of the support vector machine (SVM) by optimizing its sensitive parameters using a levy flight mutated genetic algorithm (LFMGA) to diagnose the defect in the Pelton turbine. Lou et al. [15] adopted the domain adaptation technique to diagnose the rotating components that reduce the deviation between simulated and measured signals. Qian et al. [16] also developed an improved version of SVM to diagnose autonomous vehicles. Gao et al. [17] utilized the generative adversarial network (GAN) to expand the library of faulty samples of gears to improve the identification accuracy. Vashishtha et al. [18] have used the African vulture optimization algorithm (AVOA) to optimize the parameters of time-varying filter-based empirical mode decomposition (TVF-EMD) which helps in the disintegration of the raw vibration signal into different modes for identification of rolling bearing's defects. Gao et al.[19] hybridized the FEM and GAN to diagnose the bearing defects. The FEM simulates the hard-to-get faults whereas GAN enlarges the simulated and experimental results to generate larger fault samples. Tang et al. [20] applied the Bayesian optimization (BO) for tuning hyper-parameters of the convolution neural network (CNN) for intelligent diagnosis of a hydraulic piston pump. Even though machine learning-based methods can intelligently classify the different faults, the feature extraction during the implementation of these techniques requires prior knowledge and experience. Therefore these techniques do not provide accurate results in the scenario of highly non-representable features. This necessitates the need for such techniques that have the better representation ability and requires less prior knowledge. Due to high capabilities of automated feature extraction, deep learning (DL) techniques can handle these problems easily. Kumar et al. [21] modified the objective function of the CNN by incorporating the additional sparsity in the existing objective function to diagnose the defects in the rolling bearing. Vashishtha and Kumar [22] applied generalized sparse filtering based on Wasserstein distance and maximum mean discrepancy (MMD) to diagnose the different faults in the Francis turbine and Centrifugal pump. On the basis of deep adaptation networks, a DL model was built for bearing fault diagnosis where its generalization ability was enhanced [23]. Zhao et al. [24] introduced batch normalization in the existing CNN to eliminate the feature distribution difference while diagnosing the rolling bearing under variable conditions. Zhu et al. [25] proposed the inception block and regression branch that helped in mapping the output features. Further, the positional relationship between the features classifies the bearing fault categories. Gai et al. [26] optimized the variational mode decomposition (VMD) by hybrid grey wolf optimization to decompose the signal into different modes. The sensitive features are then extracted from the prominent modes and fed into the deep belief network (DBN).

As deep learning-based methods have gained much popularity and have attained good results in fault diagnosis, but these techniques still have some limitations. For instance, (a) Deep learning-based techniques fail to encode the position and orientation of the faults and thus give inaccurate results while classifying the fault categories, and (b) The tuning of the hyperparameters of the deep learning-based techniques plays a vital role; thus they should be set intelligently [27]. The metaheuristic techniques can be used to tune the hyperparameters of the CNN. In the given work, the gorilla troop optimizer (GTO) served this task as it is easy to implement due to its simple structure. The GTO uses less storage and computational time and converges fast due to the continuous reduction of search space and fewer decision variables. Also, GTO can be applied to different engineering fields such as in conventional design problems, biomedical field, parametric selection of photovoltaic, optimal selection of parameters of fuel cells, cluster node selection in wireless sensor networks, economic load dispatch problems etc [28–31].

In this study, a worm gearbox is selected as the research test rig whose three health conditions have been analysed by the intelligent fault diagnosis scheme. In this scheme, initially, gorilla troop optimization (GTO) is enhanced by the opposition-based learning concept and quantum gate rotation concept, which is further employed to make the CNN adaptive by optimizing its hyperparameters. The continuous wavelet transform (CWT) converts the raw signals, both from vibration and acoustic, into time-frequency images by the suitable wavelet basis function. The adaptive CNN uses these images to precisely recognize the different health conditions of the worm gear system.

The rest of the manuscript is structured in the following sections. In Section 2, the background of the CNN and GTO algorithm has been discussed. In the same section, the proposed modifications, which have been incorporated into the basic GTO, are also discussed. The defect identification scheme is elaborated in Section 3. In Section 4, how defect identification is applied to the real-time application is discussed. The comparative experiments to check the efficacy of the proposed AGTO and the performance of AGTO-CNN are carried out in Section 5. The corresponding results are evaluated and discussed. The conclusions of the research work are drawn in Section 6.

**2. Related Work**

In this section, the background to the related work, such as gorilla troop optimization and CNN, has been covered.

**2.1. Convolutional Neural Network (CNN)**

CNN is a deep neural network and an example of feed-forward neural networks. It has capabilities in data mining and useful feature extraction in a supervised learning environment [32,33]. The typical structure of CNN consists of different layers as shown in Fig. 1.

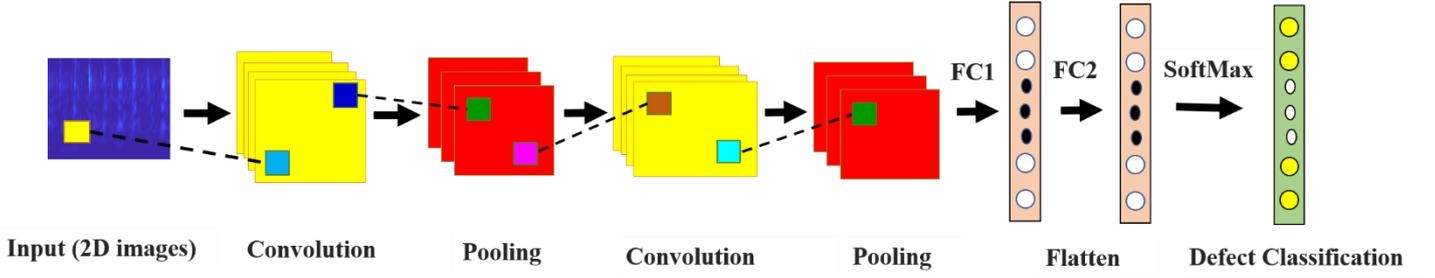

**Fig. 1:** Basis Structure of CNN

**(a) Convolution layer:** The convolution layer has a significant role in CNN for feature extraction. The filters used in CNN can cover the entire image through shifting. In this layer, the dot product is carried out between the filter and the image and then summed over the filter region [32]. The filter is then moved to the next place, covering the entire image in the same manner. The same can be expressed through Eq. (1).

$$N_v^l = F\left(\sum_{u \epsilon M_v} N_u^{l-1} * w_v^l + B_v^l\right) \quad (1)$$

where $N$ is the input of CNN, $F$ indicates the sigmoid activation function. $B_v^l$ is the bias corresponding to the $j^{th}$ output of the $l^{th}$ layer of convolutional neural network. $w_v^l$ is the weight for the $v^{th}$ feature map. $*$ represents the convolution operation and $M_v$ is used to select the input feature map.

**(b) Pooling layer:** The pooling layer performs the down-sampling operation by extracting the more significant information from the features obtained from CNN. It not only minimizes the parameters but also maintains the invariance of the image feature. The pooling operation can be understood through Eq. (2).

$$M_{v-s}^l = F\left(Q_v^l down(M_v^{l-1})\right) + B_v^l \quad (2)$$

where $F$ is an activation function, $Q_v^l$ is the weight. $B_v^l$ represent the bias and $M_v^{l-1}$ is used to denote the feature map in the $(l-1)^{th}$ layer. $down$ indicates the present pooling operation, which is performed by using average or maximum operation.

**(c) Fully connected layer:** It imitates the behaviour of ANN by connecting the neurons of one layer to other layers and converts obtained features into 2D vectors accordingly.

**(d) SoftMax layer:** This layer is part of the fully connected layer which transforms the probability distribution of target class over all classes. In this way, it maps the output value in the interval (0,1) [33]. The mathematical expression of the probability distribution in the SoftMax layer is given by Eq. (3).

$$p(y_v^l) = \frac{exp(y_v^l)}{\sum_{v=1}^{K} exp(y_v^l)} \quad (3)$$

**(e) Classification layer:** This layer calculates the loss during the training phase by minimizing the objective function. This objective function is given as

$$fit_{fn} = -\sum_{j=1}^{m} y_v^T \ln y_v^p + \varphi \sum \omega^2 \quad (4)$$

where the first term represents the cross-entropy loss function and $y^p$ is predicted value. $\varphi$ is $L2$ regularization.

### 2.2. Gorilla troop optimization (GTO)

Abdollahzadeh et al. [34] proposed the GTO by imitating the social intelligence behaviour of gorillas. The mathematical model showing the collective life of the gorillas and how they perform the exploration and exploitation is discussed in the following subsections:

#### 2.2.1. Exploration phase

The GTO algorithm imitates the behaviour of gorillas where the best gorilla at each stage of optimization becomes the candidate solution which is known as silverback. The exploration stage of optimization consists of three processes, viz., migration to an unknown location, migration towards a known location and moving to other gorillas. In GTO, these processes are achieved through the following mathematical Eq. (5).

$$GP(t+1) = \begin{cases} (X_{max} - X_{min}) \times r_1 + X_{min} & r_4 < p \\ (r_2 - C) \times X_{rand}(t) + L \times H & r_4 \geq 0.5, \\ X(t) - L \times \left(L \times \left(X(t) - GP_{rand}(t)\right) + r_3 \times \left(X(t) - GP_{rand}(t)\right)\right) & r_4 < 0.5. \end{cases} \quad (5)$$

where $r_1, r_2, r_3$ and $r_4$ are the random numbers generated within the range 0 and 1. $X_{max}$ and $X_{min}$ represents the upper and lower bounds of the variables. $GP_{rand}$ represents the position of the gorilla which is chosen randomly, and $X_{rand}$ is one of the gorillas selected randomly from the whole population. $X(t)$ is the current position of the gorilla.

$C, F, L$ and $H$ are the variables calculated as follows:

$$C = F \times \left(1 - \frac{Iter}{Max\_Iter}\right) \quad (6)$$

$$F = cos(2 \times r_5) + 1 \quad (7)$$

$$L = C \times l \quad (8)$$

$$H = Z \times X(t) \tag{9}$$

where $Iter$ and $Max\_Iter$ indicate the current and maximum iterations respectively. $r_5$ and $l$ are random numbers from [0,1], and [-1,1], respectively. $Z$ is a random variable from the range $[-C, C]$.

### 2.2.2 Exploitation stage

In this phase, the best candidate i.e. Silverback leads the whole group by taking all the decisions and directing the other gorillas towards the food source. The silverback safes and maintain the well-being of the whole group. In case, the silverback may weaken and get old or eventually die then the blackback of the group will become the group leader. But, if another male gorilla somehow engages the silverback then he will dominate the group. If $C > W$, then following the silverback mechanism is selected, but if $C < W$, adult females' competition will be preferred.

### 2.2.3. Follow the Silverback

In the group, when silverback is young and healthy. All the other gorillas follow the instructions of the silverback to visit various places in search of food. The same can be expressed by the following mathematical Eq. (10), Eq. (11) and Eq. (12).

$$GP(t+1) = L \times M \times (X(t) - X_{SB}) + X(t) \tag{10}$$

$$M = \left( \left| \frac{1}{N} \sum_{i=1}^{N} GP_i(t) \right|^g \right)^{\frac{1}{g}} \tag{11}$$

$$g = 2^L \tag{12}$$

where $X_{SB}$ represent the position of the silverback gorilla. $N$ is the total number of gorillas.

### 2.2.4. Competition for adult females

At the time of puberty, young gorillas fight with other male gorillas for female gorillas to expand their groups. These fights are violent and can last for days. The same can be expressed by the following equations.

$$GP(i) = X_{SB} - (X_{SB} \times Q - X(t) \times Q) \times A \tag{13}$$

$$Q = 2 \times r_6 - 1 \tag{14}$$

$$A = \beta \times E \tag{15}$$

$$E = \begin{cases} N_1, & r_7 \geq 0.5 \\ N_2, & r_7 < 0.5 \end{cases} \tag{16}$$

where $r_6$ and $r_7$ are the random numbers within range (0,1). $E$ is a random number from the normal distribution.

### 2.3. Proposed modifications in GTO

In this section, the proposed modifications which enhance the searchability of the basic GTO in the basic GTO have been discussed.

### 2.3.1. Opposition-based learning

Most of the metaheuristics optimization algorithms initiate with random global solutions which initializes the individual in the given search space. Further, each individual updates their position in the search for appropriate solutions on the basis of their intellect. This process is time-consuming which can be addressed by improving the search process and acquire better results by simultaneously using the opposite solution of initial guesses. So the optimization algorithm has the choice to select either from a random solution or its opposite guess. As a result, a decision can be made by the optimization algorithm that can speed up convergence. Not only the same procedure can be applied to beginning positions, but it may also be applied to each position in the current population [18,35]. The initialization process in the opposition-based learning concept is shown in Eq. (17).

$$X_{i+N,j}^t = X_{min_j} + X_{max_j} - X_{i,j}^t \tag{17}$$

where $(i = 1,2,\ldots,N; j = 1,2,\ldots,D)$

### 2.3.2. Concept of the Quantum Gate Rotation (QRG)

The QRG is a state processing method used in the field of quantum computing. The position data produced by the swarm-based method is floating-point data, whereas quantum bits are binary. The discrete quantum bit data must be converted into continuous algorithm data to process the location information [36]. A quantum rotation gate rotates the data of each dimension of the search agent in pairs and updates it. Eqs. (18) and (19) show the updation and adjustment procedures for QRG. The same can be observed in Fig. 2.

$$\mu(\theta_i) = \begin{bmatrix} cos(\theta_i) & -sin(\theta_i) \\ sin(\theta_i) & cos(\theta_i) \end{bmatrix} \tag{18}$$

$$\begin{bmatrix} a_i' \\ b_i' \end{bmatrix} = \mu(\theta_i) \begin{bmatrix} a_i \\ b_i \end{bmatrix} = \begin{bmatrix} cos(\theta_i) & -sin(\theta_i) \\ sin(\theta_i) & cos(\theta_i) \end{bmatrix} \begin{bmatrix} a_i \\ b_i \end{bmatrix} \tag{19}$$

where $[a_i \quad b_i]^T$ is the state of the $i^{th}$ quantum bit before applying QRG. Whereas $[a_i' \quad b_i']^T$ indicate the state for the same quantum after QRG. $\theta_i$ is the rotation angle for $i^{th}$ quantum bit and calculated through Eq. (20)

$$\theta_i = \Delta\theta_i \times r(a_i, b_i) \tag{20}$$

where $r(a_i, b_i)$ indicates rotation in the target direction. $\Delta\theta_i$ indicates the rotation angle of the $i^{th}$ rotation. The direction of the target having the highest fitness is chosen to rotate the individual for exploring the search field by comparing the fitness value of the current target with the optimal target.

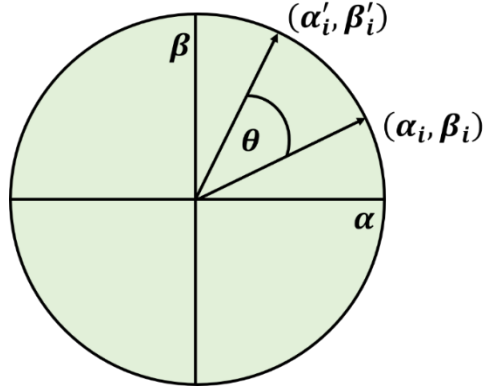

**Fig. 2** Quantum gate rotation

The rotation angle of the QRG is chosen to maintain the balance between exploration and exploitation. When the current individual is far from the best, the value of $\theta$ should be raised during the early exploration stage whereas it should be lowered during the exploitation phase [36]. This approach makes it possible for the search process to adapt to various solutions and is better suited for searching for the overall best solution. The rotation angle is given as

$$\Delta\theta = \theta_{min} + \gamma_i(\theta_{max} - \theta_{min}) \tag{21}$$

where $\theta_{max}$ and $\theta_{min}$ are the maximum and minimum values of $\Delta\theta$ which are $0.035\pi$ and $0.001\pi$, respectively. $\gamma$ is a function that can be calculated as follows:

$$\gamma_i = 1 - e^{-4.\left(\frac{bF-X(i)}{bF-wF}\right)^2} \tag{22}$$

where $bF$ and $wF$ are the best and worst fitness functions of the current iteration.

The pseudo-code of the proposed optimization algorithm is given in Algorithm 1.

**Algorithm 1: Pseudo code of the proposed algorithm**

```
Input: Population size (N), maximum number of iterations, (Max_Iter), controlling parameters p, β, and W
Output: Best fitness
Initialization of the random population (X)
Apply opposition − based learning on the initial population (X) and obtain (2N) population
Evalaute the fitness function of each gorilla (2N)
Sort the fitness vaules and save best (N)population
For iter = 1: Max_Iter
Update the Value of C and L using Eq. (6)and (8)
        For i = 1: N
        Update the position of gorilla using Eq. (5)
        End For
        Calculate the fitness of each gorilla
        if GP is better than X, replace them
```

$$\begin{aligned}
&\quad\quad\quad\text{save the best location as } X_{SB}\\
&\quad\quad\quad\textbf{For } i = 1:N\\
&\quad\quad\quad\quad\quad\textbf{If } |C| \geq 1\\
&\quad\quad\quad\quad\quad Update\ the\ position\ of\ gorilla\ using\ Eq.(10)\\
&\quad\quad\quad\quad\quad\textbf{Else}\\
&\quad\quad\quad\quad\quad Update\ the\ position\ of\ gorilla\ using\ Eq.(13)\\
&\quad\quad\quad\quad\quad\textbf{End If}\\
&\quad\quad\quad\textbf{End For}\\
&\quad\quad\quad Calculate\ the\ fitness\ of\ each\ gorilla\\
&\quad\quad\quad if\ new\ solution\ show\ improvment\ over\ previous\ solutions, replace\ them\\
&\quad\quad\quad save\ the\ best\ location\ as\ X_{SB}\\
&\quad\quad\quad Evalaute\ the\ value\ of\ \gamma_i\ using\ Eq.(22)\\
&\quad\quad\quad Calculate\ the\ rotation\ angle\ \theta_i\ using\ Eq.(20)\\
&\quad\quad\quad Apply\ the\ quantum\ gate\ rotation\ on\ each\ gorilla\\
&\quad\quad\quad Create\ a\ pool\ of\ gorillas\ and\ Y, and\ evaluate\ fitness\ of\ each\ member\ in\ this\ pool\\
&\quad\quad\quad\ Sort\ the\ best\ N\ positions\ of\ gorilla\\
&\quad\quad\quad\textbf{End For}\\
&\quad\quad\quad\ \textbf{Return}\ bestfitness
\end{aligned}$$

## 3. Fault diagnosis scheme

An amended Gorilla troop optimization (AGTO) algorithm is built to optimize the different parameters of CNN, such as learning rate, epoch, batch size, activation and number of neurons in hidden layers, considering the loss(error) as cost function as given in Eq. (4). AGTO minimizes the fitness function. The process of optimizing the hyperparameters of CNN by amended GTO is shown in Fig. 3.

The methodology adapted to diagnose the worm gearbox defects is shown in Fig. 4. This methodology can be implemented by following steps:

(1) Acquisition of data through a uniaxial accelerometer and an acoustic sensor.

(2) The raw signals are converted into time-frequency images by the CWT. The Morlet wavelet function is used to serve this purpose. The 2D images were processed by the transform strategy and further utilized to feed the CNN.

(3) Initially, the CNN model was developed with random values of its hyperparameters.

(4) Then, the CNN model is optimized by the proposed AGTO algorithm. The hyperparameters which have been optimized are the number of neurons in hidden layers, learning rate, batch size, epoch and activation function. The range for each hyperparameter is selected from the literature [37–39]. The main purpose of the optimization is to choose the set of optimal hyperparameters which enhances the performance of the CNN with less computational complexity.

(5) At the optimal combination of hyperparameters, the CNN with AGTO is used to recognize the different faults of the worm gearbox.

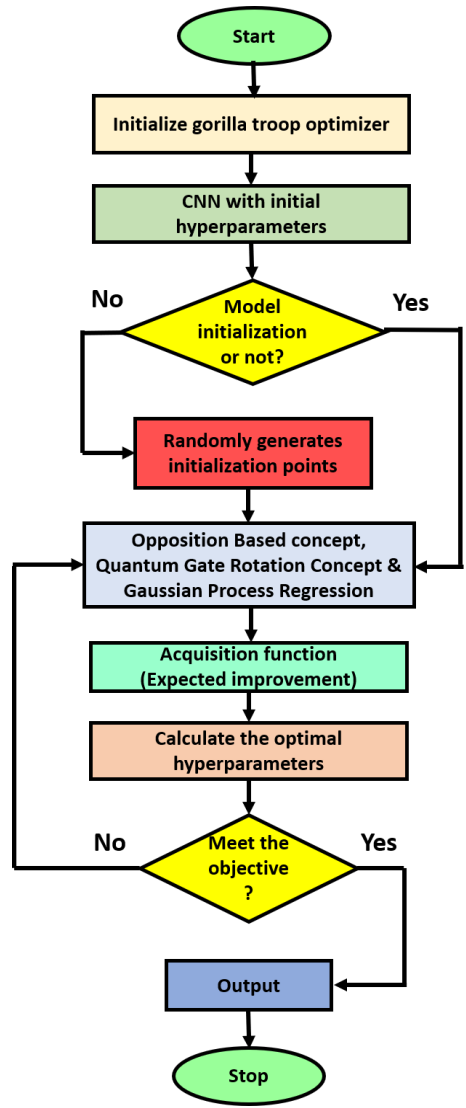

**Fig. 3** Flow chart showing optimization of HPs of CNN by AGTO

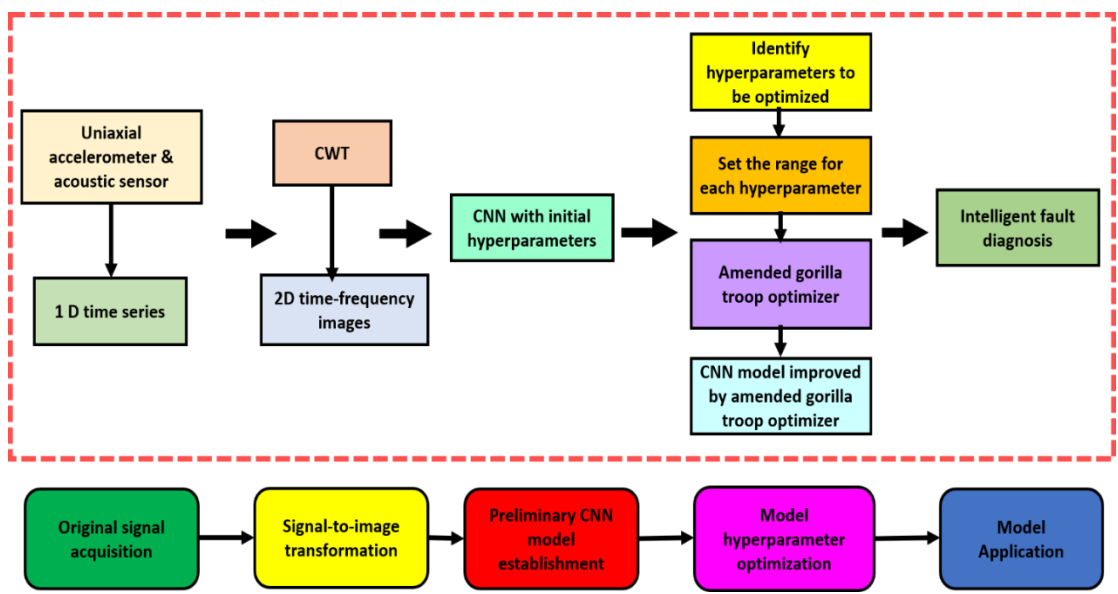

**Fig. 4** Fault diagnosis of worm gearbox through AGTO-CNN model

## 4. Application of fault diagnosis scheme to worm gearbox

## 4.1 Test Rig

The proposed diagnosis scheme is applied on the test rig of the worm gearbox. The worm gearbox is driven by the 50 Hz DC motor connected through it by a flexible coupling. The DC motor is controlled by the control panel. A schematic and picture of the test rig is shown in Fig. 5.

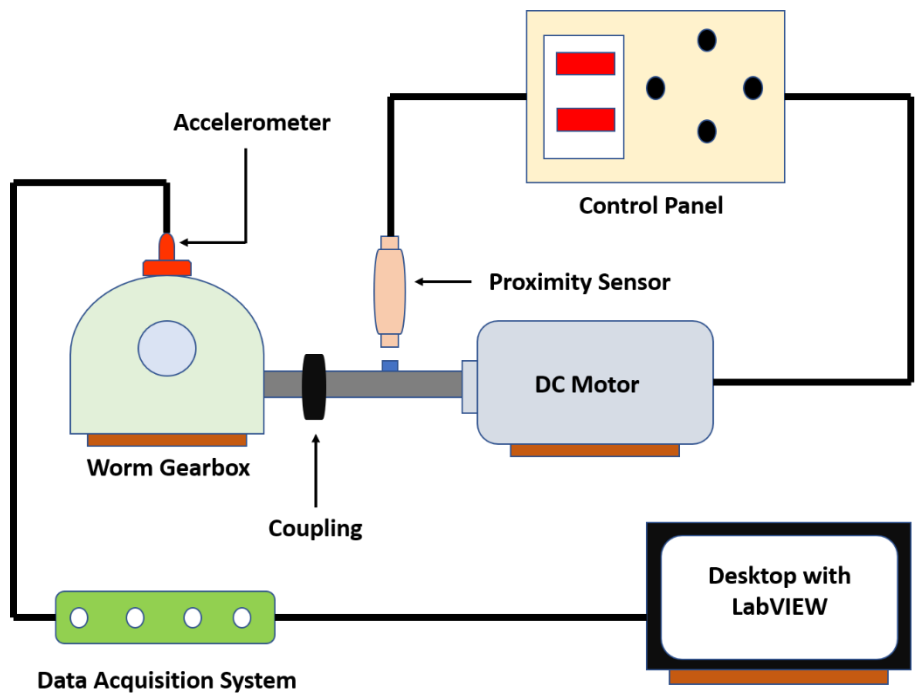

(a)

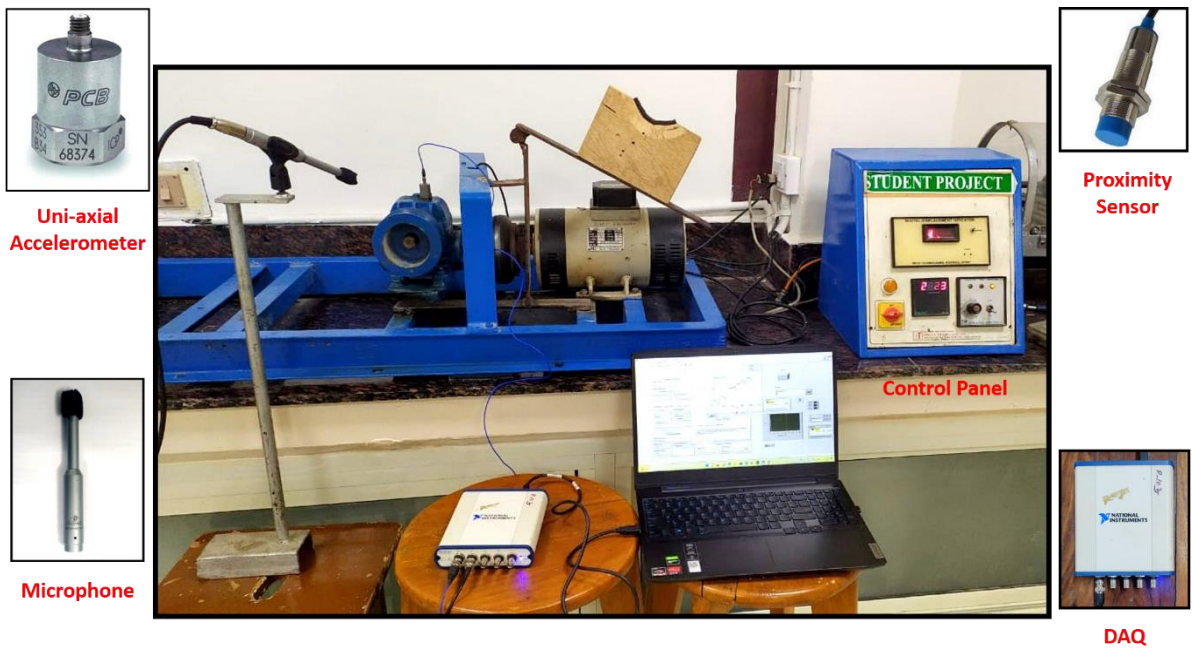

(b)

**Fig. 5** Worm gearbox test rig (a) schematic view and (b) pictorial view

## 4.2. Data Acquisition

The vibration and acoustic signals are acquired from the worm gearbox under three different health conditions at rated rpm of 910, 1500 and 2520. The three different health conditions which are taken into consideration are healthy (no defect), pitting and missing, as shown in Fig.6.

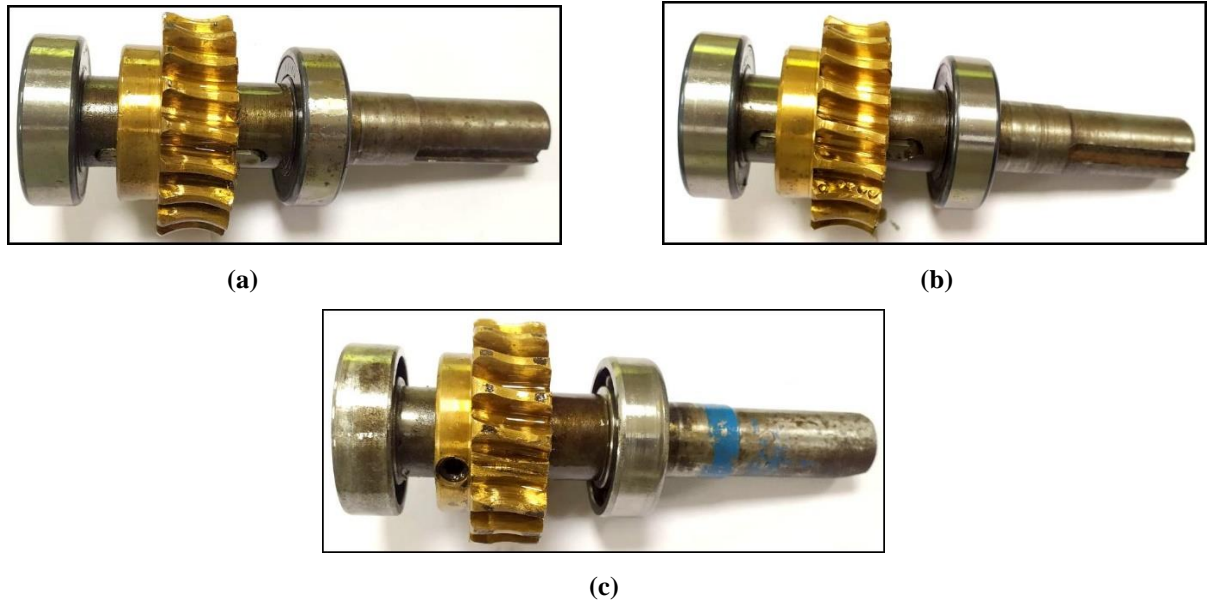

**Fig. 6** Health states of worm gearbox (a) Healthy (b) Pitting and (c) Missing

Initially, the worm is defect free and in place; this condition is considered to be healthy. But chances of some inherent defects may be there. For each health condition, a total of 900 signals are acquired at the rated rpm of 910, 1500 and 2520. The data is acquired with the help of PCB® Piezotronics make uniaxial accelerometer having a sensitivity of 100 mV/g which was mounted on the gearbox. Whereas the sound signals/ acoustic signals were recorded by the microphone of ECM 8000 make, having a sensitivity of -60 dB. The DAQ with 24-bit and 4-channel of National Instrument make is used to acquire vibration signal in LabView environment of 2020 version. The analysis has been done on MATLAB R2019a software. The configurations of the machine are AMD Ryzen 5 4600 H with radion graphics 3 GHz with 8 GB of RAM and 64-bit Windows 10 operating system.

## 4.3. Data preprocessing

The raw vibration signal and acoustic signals obtained from the test rig for the different health conditions of the worm gearbox are shown in Fig. 7.

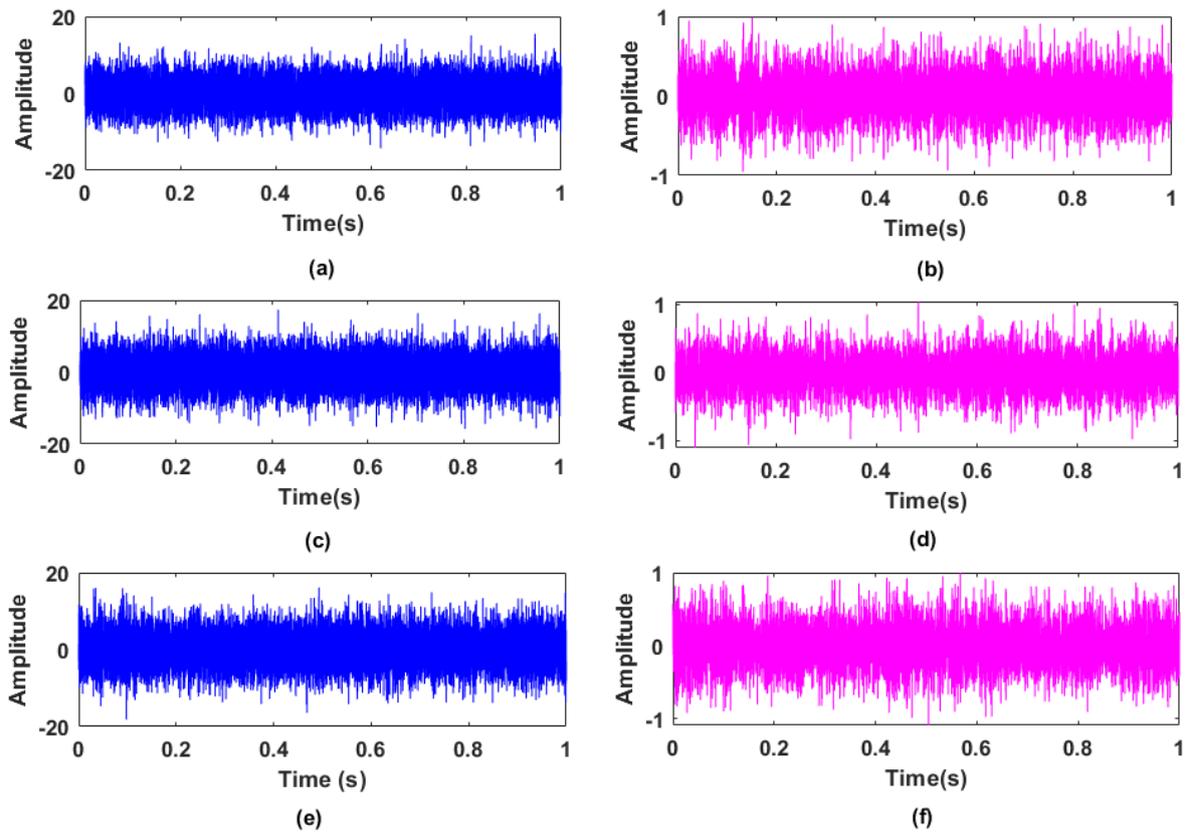

**Fig. 7** Raw signals (a) vibration signal of healthy, (b) acoustic signal of healthy, (c) vibration signal of pitting (d) acoustic signal of pitting (e) vibration signal of missing and (f) acoustic signal of missing

These signals are in 1D time series, which must be transformed into 2D because only 2D images can be fed to CNN. The acquired signal may be submerged by the background noise, or it can be affected by the long transmission lines. Therefore, researchers have used various pre-processing techniques such as the short-time Fourier transform, wavelet synchro-squeezed transform and so on to carry out the time-frequency analysis. The signals obtained from the worm wheel test rig are non-stationary ones. And for non-stationary signals, CWT is preferred as it not only extracts the time-frequency features from the original signals but can map 1D signal into 2D space. Researchers have suggested that non-orthogonal wavelet functions are preferred for getting smooth and continuous wavelet amplitudes. Thus, the Morlet wavelet function has been selected for converting the vibration signals into images as it not only has non-orthogonality but also has a good trade-off between time and frequency. The obtained images for different health states are shown in Fig. 8.

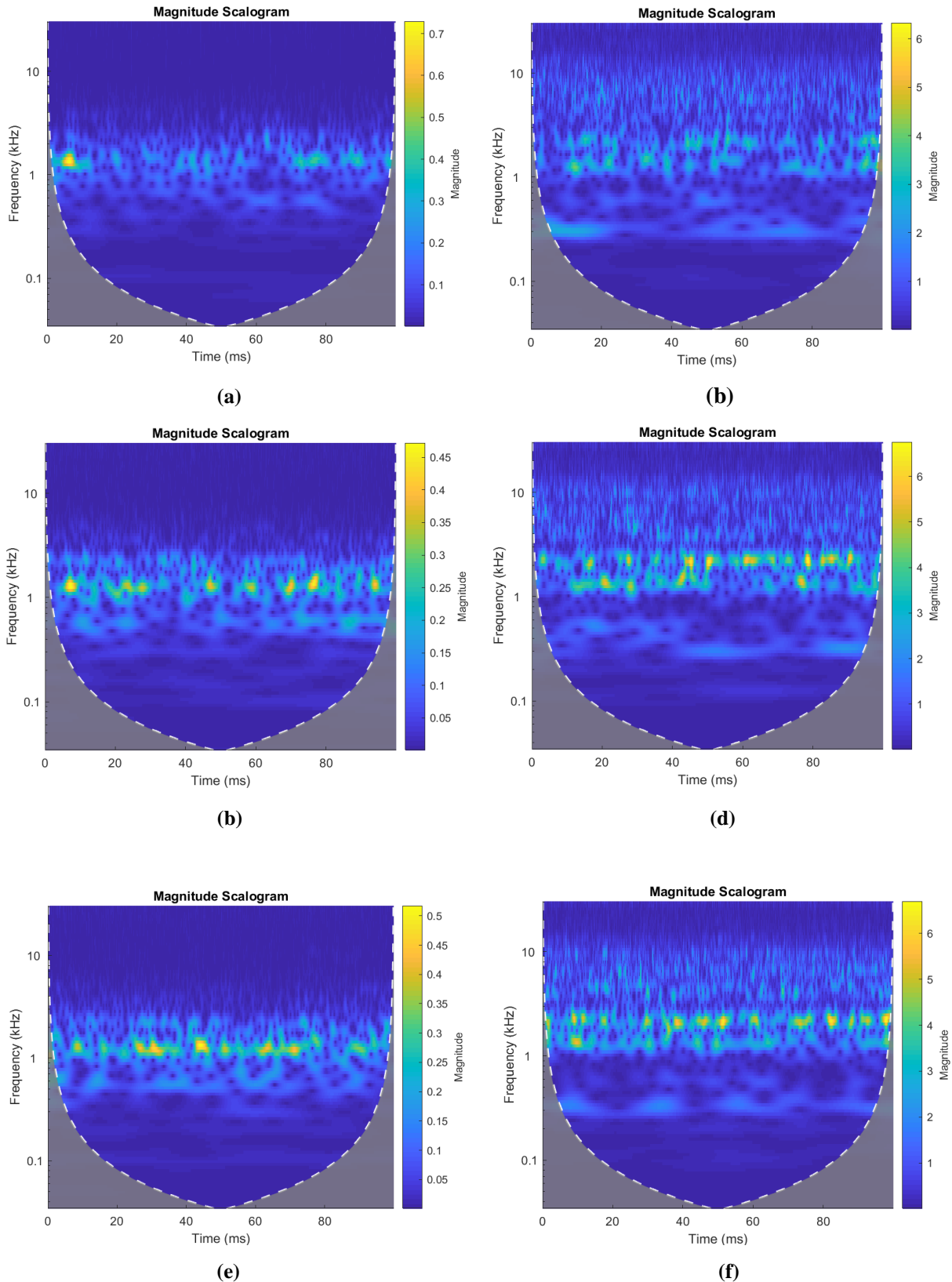

**Fig. 8** Time-frequency images from (a) vibration signal of healthy, (b) acoustic signal of healthy, (c) vibration signal of pitting (d) acoustic signal of pitting (e) vibration signal of missing and (f) acoustic signal of missing

It is obvious from Fig. 8 that the time-frequency images have a strong similarity which makes it difficult to detect any difference between the images. Thus it becomes very difficult to detect any useful information from such similar images. Therefore it is important to develop an intelligent and robust technique that can learn prominent features and recognise the types of defects simultaneously.

From three different health conditions of the worm gear, a total of 900 vibration and sound signals have been acquired. From these 900 signals, a total of 8400 time-frequency images have been extracted, i.e. 2800 for each health condition to make the dataset. Further, the dataset has been divided into training and testing datasets in the ratio of 8:2. This means the training dataset has 6720 images, whereas the testing dataset has 1680 images which have been used to train and test the built CNN model. The training samples (images) have been randomized in order to allow more prominent features to be extracted.

## 5. Result and Discussion

### 5.1. Comparison of AGTO with other algorithms

The performance of the proposed AGTO algorithm was investigated on classical benchmark functions (defined in Table 1) in terms of average (Avg) and standard deviation (Std). The results of the AGTO algorithm on classical benchmark functions have been ranked through the Friedman test to draw more reasonable conclusions. The benchmark functions consist of unimodal, multimodal, and fixed-dimension multimodal functions. The unimodal functions (F1-F7) are typically utilized to assess the algorithm's local exploitation potential because they have only one local solution and one optimal global solution. Multimodal functions (F8–F13) are frequently employed to evaluate the algorithm's capacity for exploration. To assess the stability of the approach, F14–F23, which are fixed-dimensional multimodal functions with lots of local optimal points and low dimensionality, can be employed. Wilcoxon's rank-sum test has also been carried out to evaluate the statistical performance of the proposed AGTO algorithm.

**Table 1**
Definition of classical benchmark functions

| Function type | Function | Name of function | Search Range | Dimension | Global optimum |
|---|---|---|---|---|---|
| **Unimodal functions** | F1 | Sphere | [-100,100] | 30 | 0 |
| | F2 | Schwefel 2.22 | [-10,10] | 30 | 0 |
| | F3 | Schwefel 1.2 | [-100,100] | 30 | 0 |
| | F4 | Schwefel 2.21 | [-100,100] | 30 | 0 |
| | F5 | Rosenbrock | [-30,30] | 30 | 0 |
| | F6 | Step | [-100,100] | 30 | 0 |
| | F7 | Quartic | [-1.28,1.28] | 30 | 0 |
| **Multimodal functions** | F8 | Schwefel | [-500,500] | 30 | -418.9829*D |
| | F9 | Rastrigin | [-5.12,5.12] | 30 | 0 |
| | F10 | Ackley | [-32,32] | 30 | 0 |
| | F11 | Griewank | [-600,600] | 30 | 0 |
| | F12 | Penalized | [-50,50] | 30 | 0 |
| | F13 | Penalize 2 | [-50,50] | 30 | 0 |

| | F14 | Foxholes | [-65,65] | 2 | 0.998004 |
|---|---|---|---|---|---|
| | F15 | Kowalik | [-5,5] | 4 | 0.0003075 |
| Fixed-dimensional multimodal functions | F16 | Six-hump Camel-Back | [-5,5] | 2 | -1.03163 |
| | F17 | Branin | [-5,5] | 2 | 0.398 |
| | F18 | Goldstein-Price | [-2,2] | 2 | 3 |
| | F19 | Hartman3 | [-1,2] | 3 | -3.8628 |
| | F20 | Hartman6 | [0,1] | 6 | -3.322 |
| | F21 | Shekel 5 | [0,10] | 4 | -10.1532 |
| | F22 | Shekel 7 | [0,10] | 4 | -10.4028 |
| | F23 | Shekel 10 | [0,10] | 4 | -10.5363 |

The AGTO has been compared to basic GTO and eight other optimization algorithms such as arithmetic optimization algorithm (AOA), dragonfly algorithm (DA), grey wolf optimizer (GWO), moth flame optimizer (MFO), multiverse optimizer (MVO), sine cosine algorithm (SCA), and salp swarm algorithm (SSA). The parameter setting of each algorithm is tabulated in Table 2. For a fair comparison, the population size and maximum number of function evaluations are set to 30 and 15,000, respectively. Because of the stochastic nature of the metaheuristic algorithm, the outcome may differ each time it is run. Therefore, the algorithm is performed 30 times autonomously to acquire the global solution.

**Table 2**
Parametric settings of different algorithms

| Algorithm | Parameter | Value |
|---|---|---|
| AGTO | $\beta$ | 3 |
| | $W$ | 0.8 |
| | $p$ | 0.03 |
| AOA | $r_1$ | $random$ |
| | $r_2$ | $random$ |
| | $r_3$ | $random$ |
| | $\alpha$ | 5 |
| | $\mu$ | 0.499 |
| ALO | $a$ | $min(all\ varibles\ at\ l^{th}\ iteration)$ |
| | $b$ | $max(all\ varibles\ at\ l^{th}\ iteration)$ |
| | $I$ | $10^w \left(\frac{l}{L}\right)$ |
| DA | $s$ | $2 * random * mc$ |
| | $a$ | $2 * random * mc$ |
| | $c$ | $2 * random * mc$ |
| | $f$ | $2 * random$ |
| | $e$ | $mc$ |
| | $mc$ | $0.1 - l * \left(\frac{0.1}{(L/2)}\right)$ |
| GWO | $convergence\ parameter\ (a)$ | $Linear\ reduction\ from\ 2\ to\ 0$ |
| MFO | $Convergence\ constant\ (a)$ | [-2-1] |
| | $spiral\ factor\ (b)$ | 1 |
| MVO | $r_1$ | $random$ |
| | $r_2$ | $random$ |
| | $r_3$ | $random$ |
| | $p$ | 6 |
| | $c_1$ | $2 * e^{-\left(\frac{4l}{L}\right)^2}$ |
| SSA | $c_2$ | $random$ |
| | $c_3$ | $random$ |
| SCA | $a$ | 2 |
| | $r_1$ | $a - l\left(\dfrac{a}{L}\right)$ |
| | $r_2$ | $2*\pi*random$ |
| | $r_3$ | $2*random$ |
| | $r_4$ | $random$ |

### 5.1.1. Qualitative results on 23 classical functions

The qualitative performance of AGTO is examined through the convergence curve and average fitness of the population. The convergence curve indicates the fitness value of the silverback as the best solution during the optimization process. In contrast, the average fitness curve shows how the average fitness of the whole population changes in various optimization stages. It can be observed from Fig. 9 that gorillas have initiated the exploitation operation in the close proximity to the optimal solution and then continued the exploration operation in the search space. It is clear from the convergence curve that AGTO converges rapidly and has a great capacity to enhance all gorillas in at least half of the iterations.

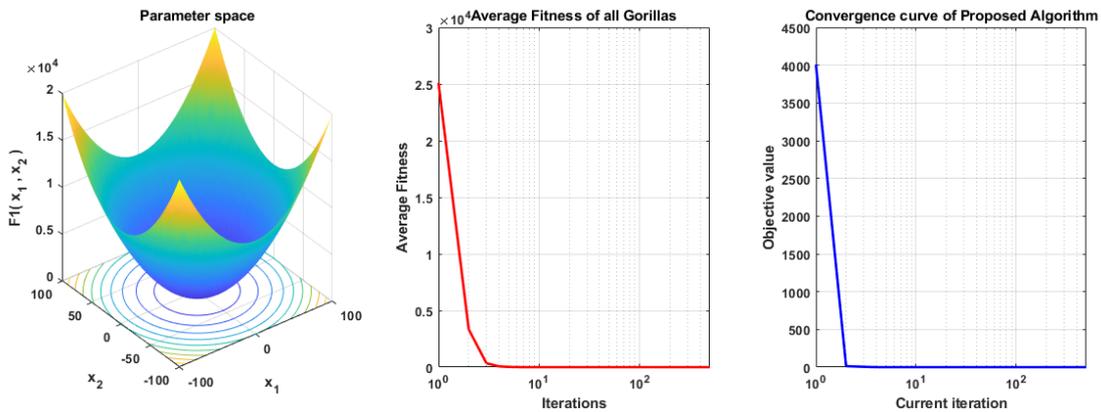

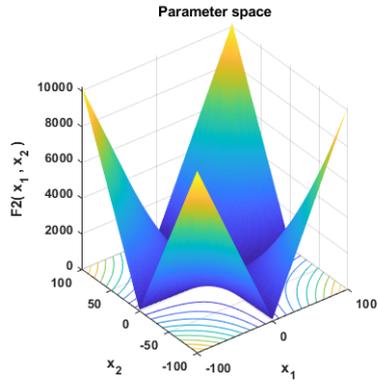
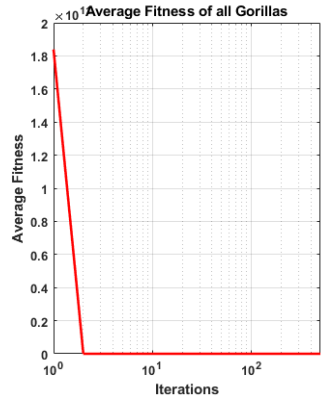
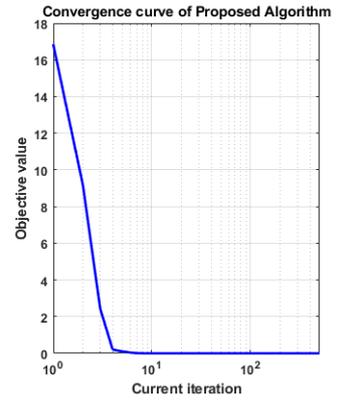
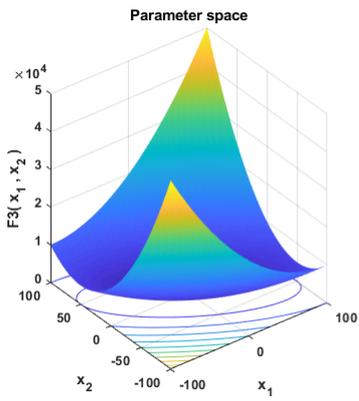
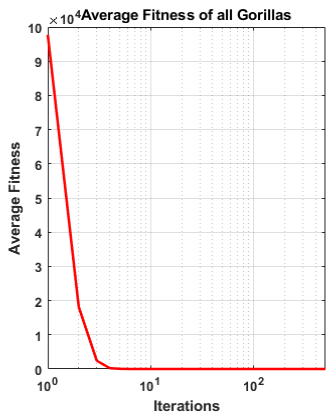
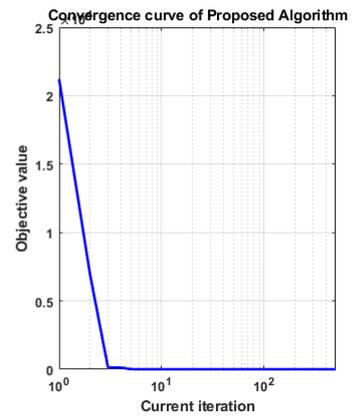
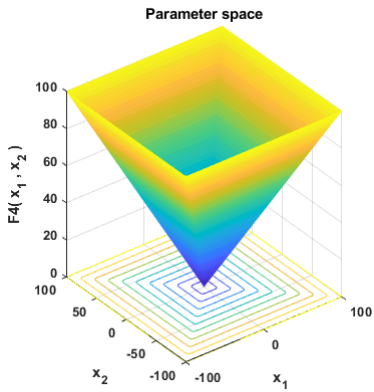
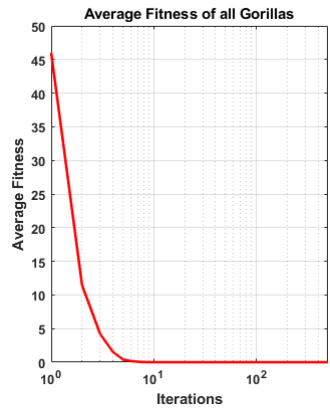
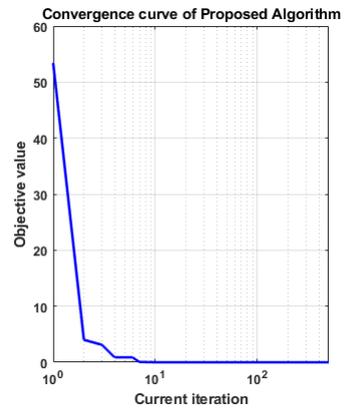
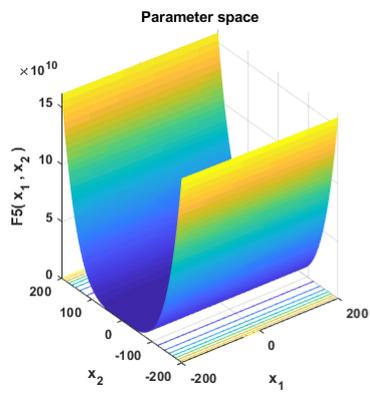
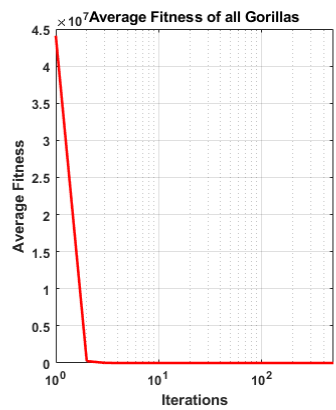
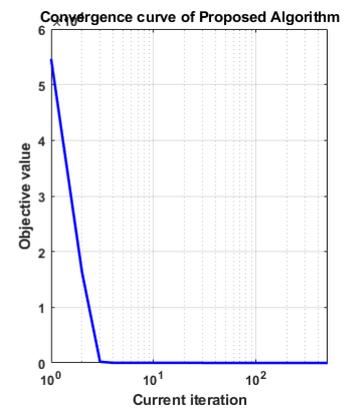

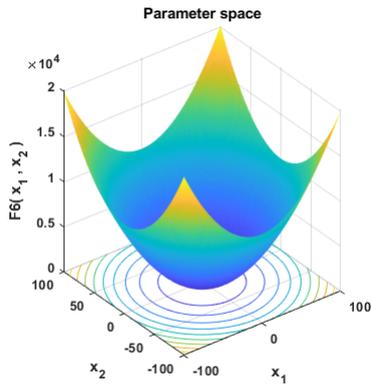
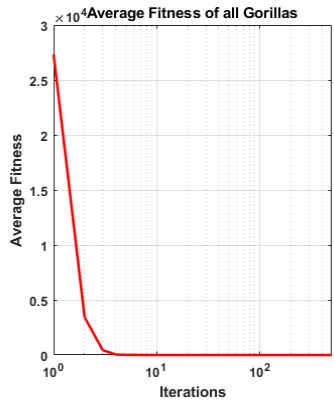
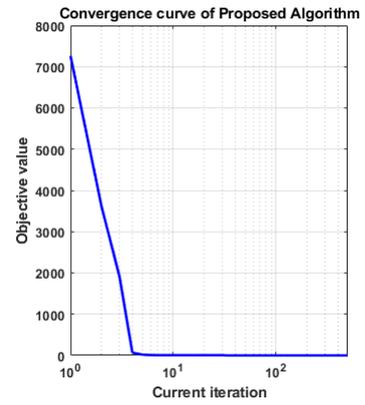
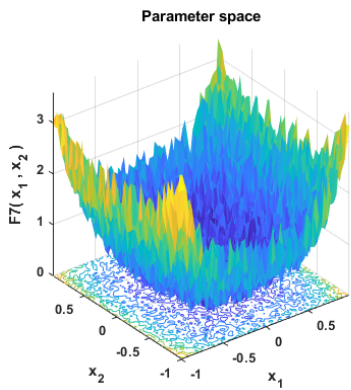
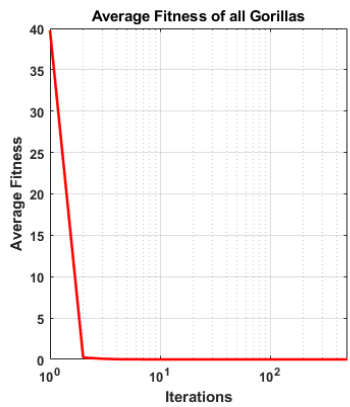
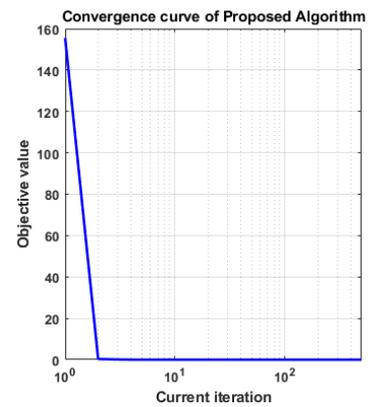
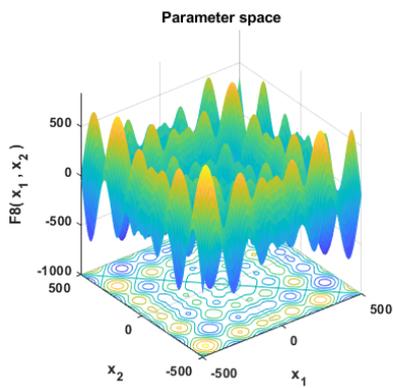
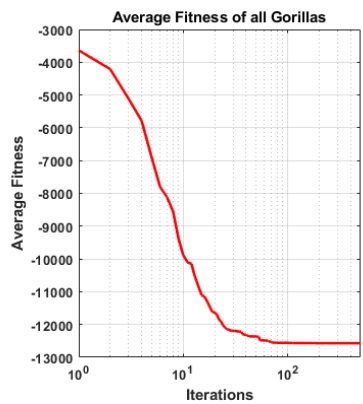
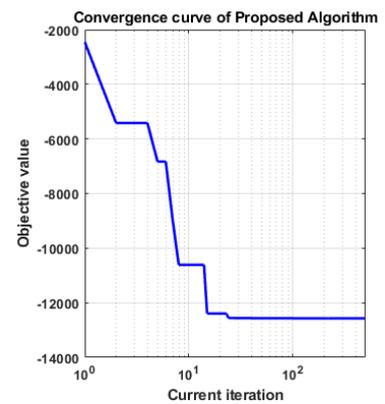
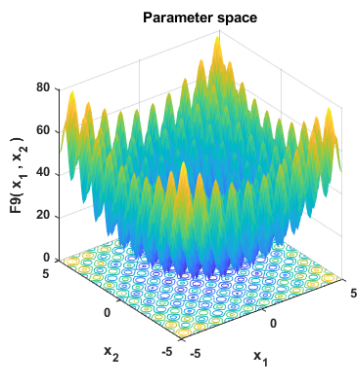
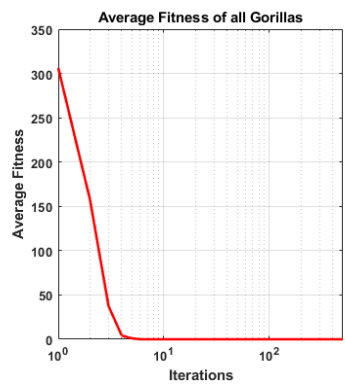
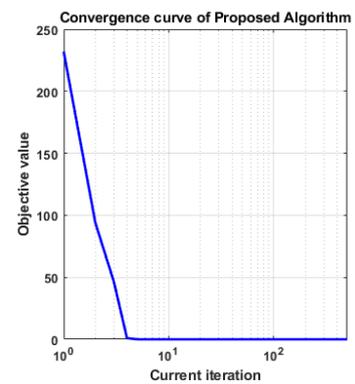

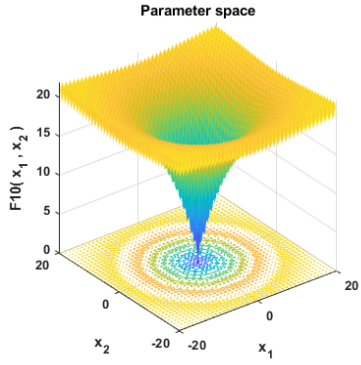 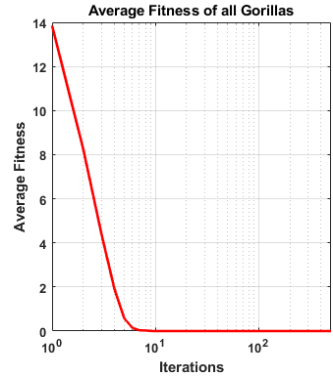 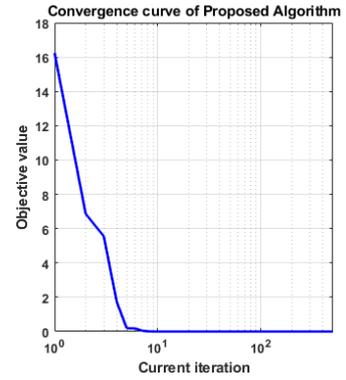
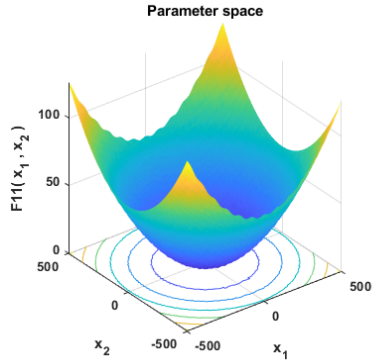 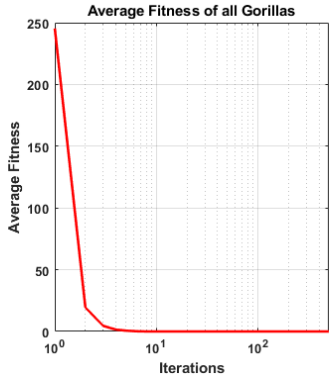 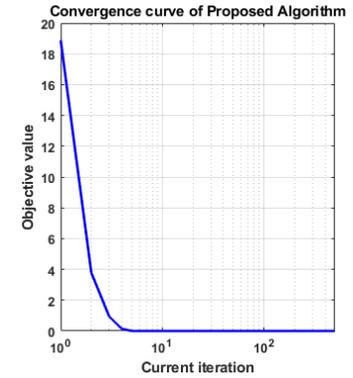
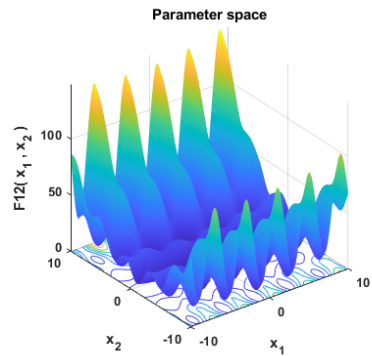 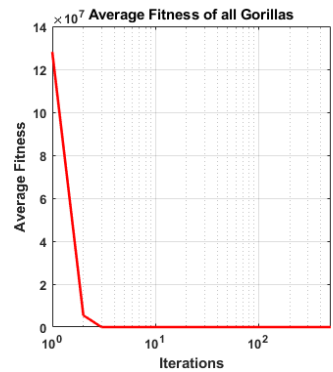 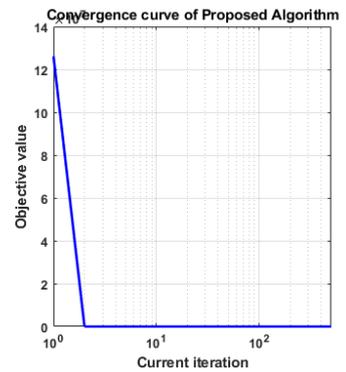
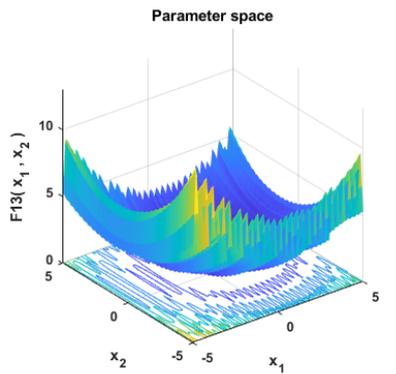 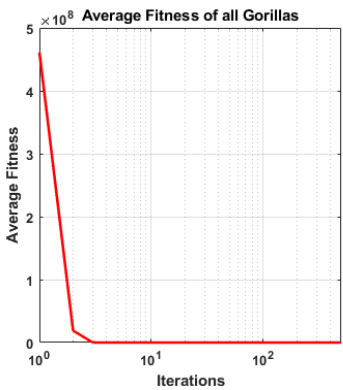 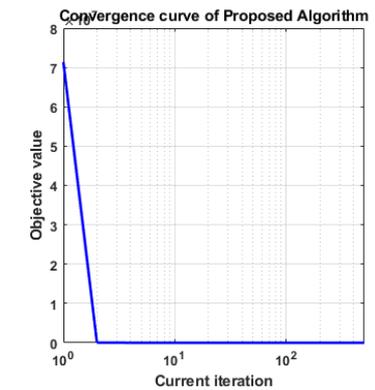

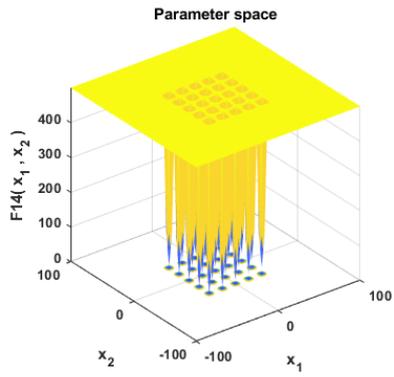
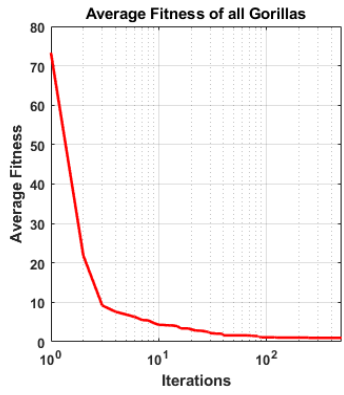
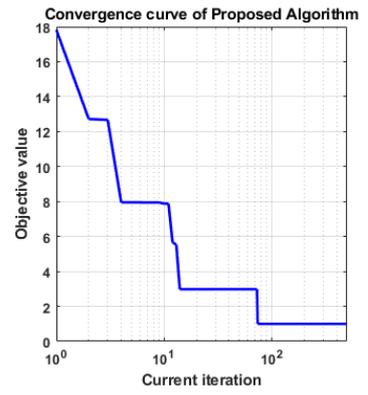
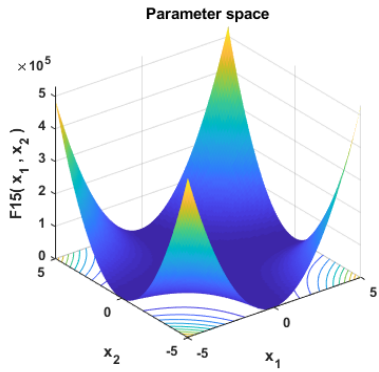
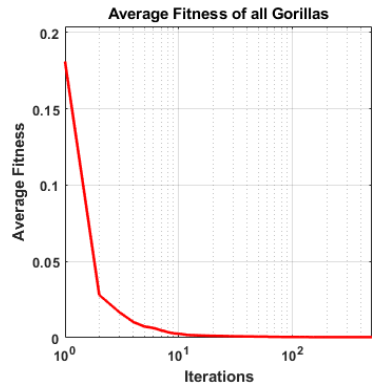
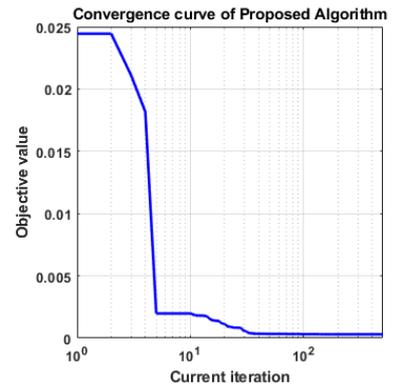
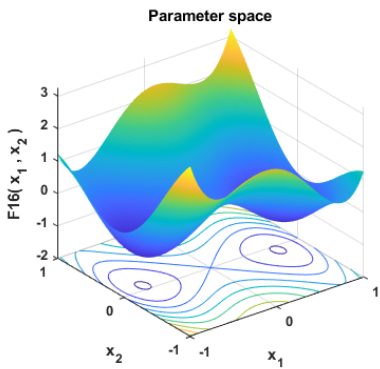
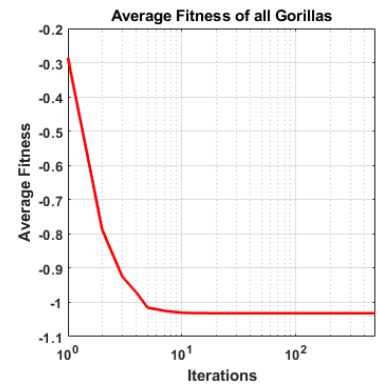
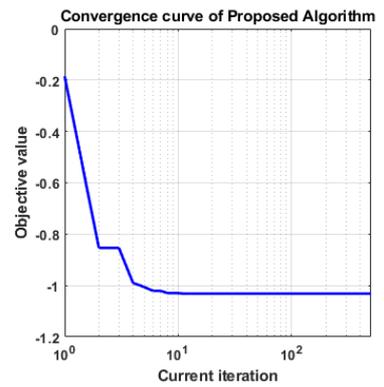
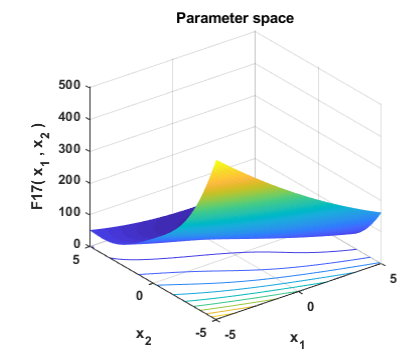
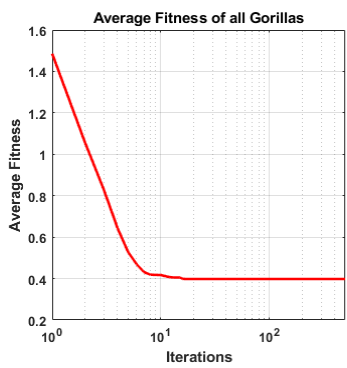
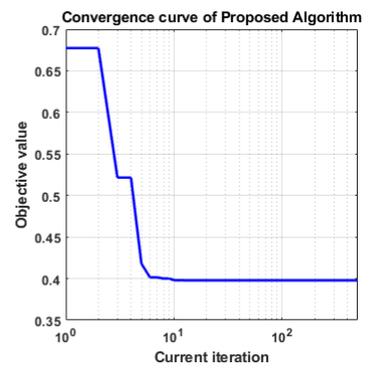

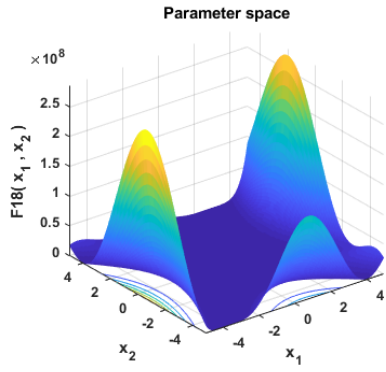
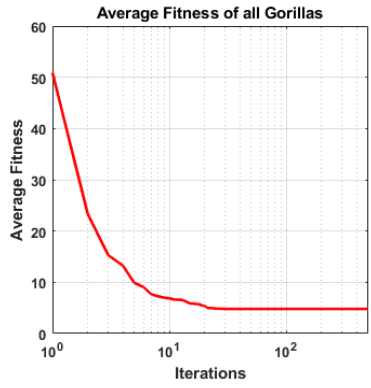
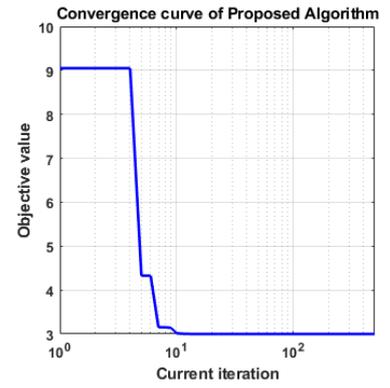
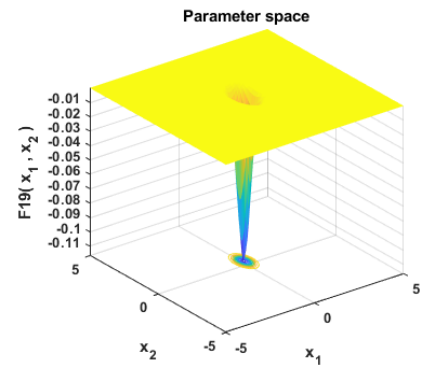
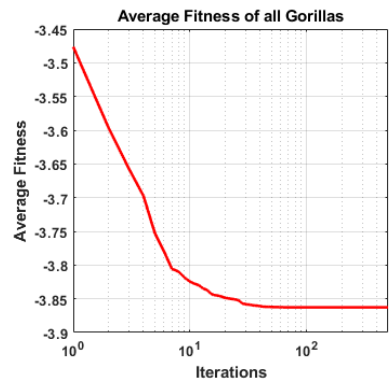
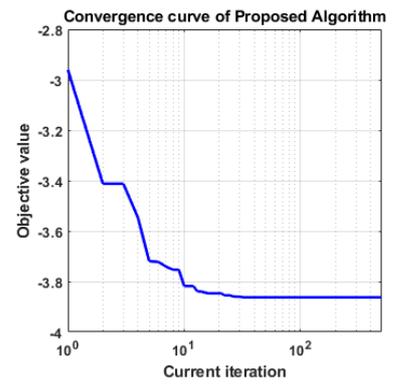
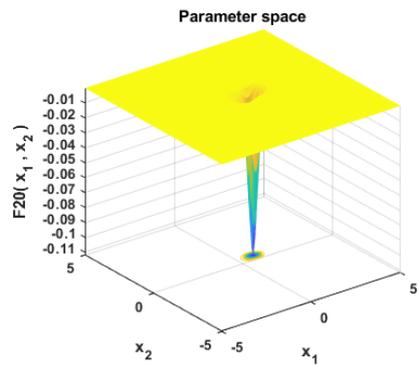
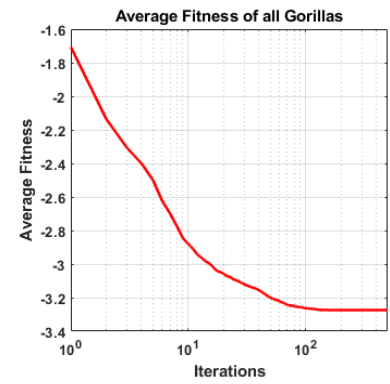
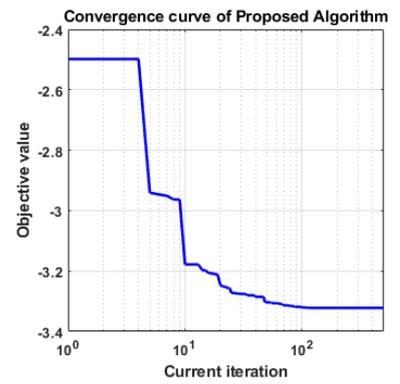
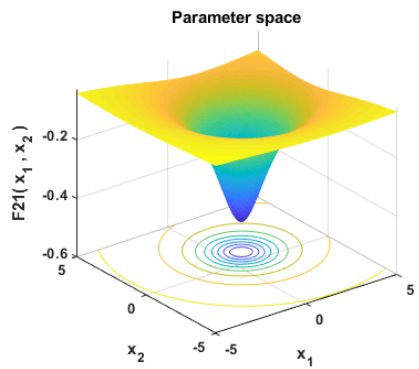
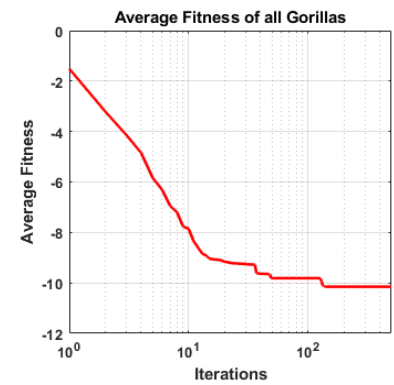
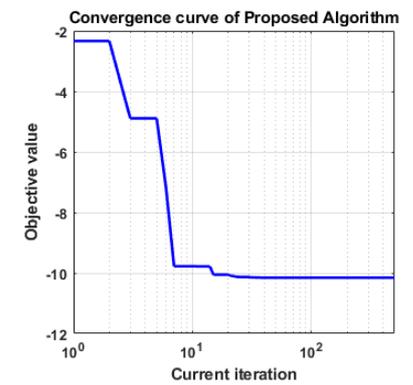

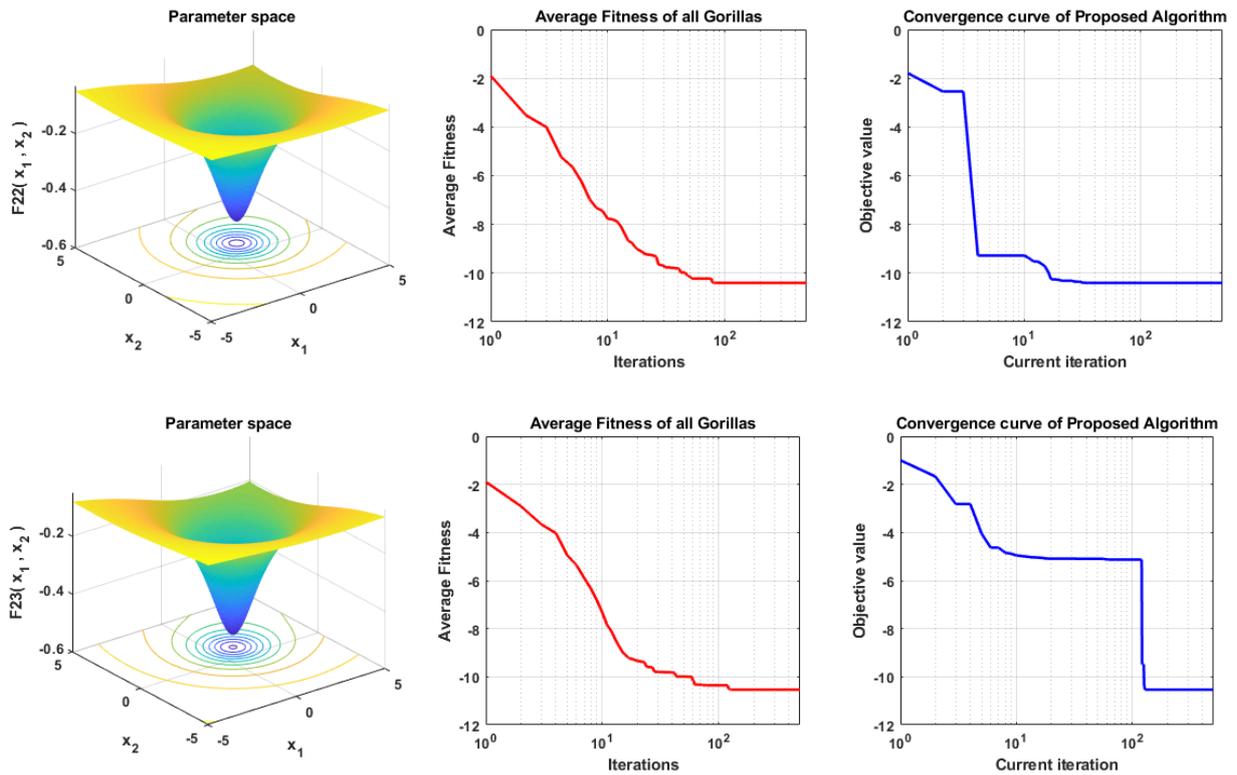

**Fig. 9** Qualitative results of the proposed AGTO algorithm

The convergence behaviour of the proposed AGTO algorithm has also been compared to the traditional GTO algorithm as shown in Fig. 10. It is clear from Fig. 10, that the AGTO converges at a faster rate than that of traditional GTO.

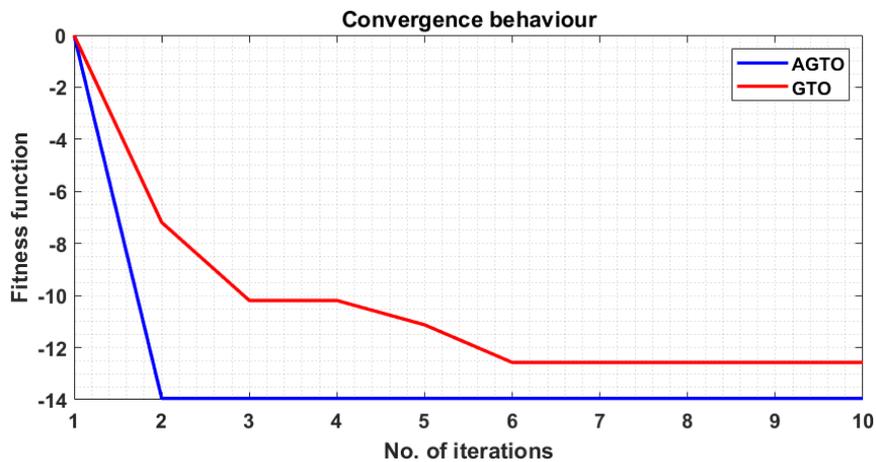

**Fig. 10** Convergence behaviour of AGTO and traditional GTO

### 5.1.2. Quantitative results on 23 classical functions

The quantitative performance of the proposed AGTO is examined on the benchmark functions in terms of Avg and Std. The results of the proposed AGTO algorithm have been compared to validate the efficacy of the AGTO. The results of the AGTO algorithm on classical benchmark functions have been ranked through the Friedman test to draw more reasonable conclusions. The obtained results are tabulated in Table 3.

**Table 3**
Results of the proposed algorithm at benchmark functions

|     |      | AOA      | ALO      | DA       | GWO      | MFO      | MVO      | SCA      | SSA      | GTO       | AGTO     |
| --- | ---- | -------- | -------- | -------- | -------- | -------- | -------- | -------- | -------- | --------- | -------- |
| F1  | Avg  | 3.73E-64 | 7.03E-06 | 1478.518 | 1.13E-16 | 1666.667 | 0.285773 | 0.333526 | 1.29E-08 | 0         | 0        |
|     | Std  | 2.04E-63 | 6.22E-06 | 766.2623 | 4.96E-17 | 3790.49  | 0.095841 | 1.714933 | 3.65E-09 | 0         | 0        |
|     | Rank | 2        | 5        | 8        | 3        | 9        | 6        | 7        | 4        | 1         | 1        |
| F2  | Avg  | 0        | 35.34462 | 13.55902 | 5.39E-08 | 40.33356 | 0.43547  | 3.96E-05 | 1.114872 | 1.68E-192 | 0        |
|     | Std  | 0        | 46.88196 | 5.043681 | 1.84E-08 | 24.70265 | 0.13303  | 7.20E-05 | 1.260599 | 0         | 0        |
|     | Rank | 1        | 8        | 7        | 3        | 9        | 5        | 4        | 6        | 2         | 1        |
| F3  | Avg  | 0.002365 | 1151.9   | 15356.97 | 443.0136 | 16766.67 | 46.24936 | 4044.797 | 368.6763 | 0         | 0        |
|     | Std  | 0.00881  | 666.1661 | 10095.45 | 188.0379 | 11242.6  | 18.8267  | 3880.241 | 254.8877 | 0         | 0        |
|     | Rank | 2        | 6        | 8        | 5        | 9        | 3        | 7        | 4        | 1         | 1        |
| F4  | Avg  | 0.021292 | 11.55389 | 23.06015 | 1.104124 | 66.64077 | 1.086291 | 18.0148  | 8.17505  | 1.16E-192 | 0        |
|     | Std  | 0.019539 | 3.317957 | 6.965639 | 1.064798 | 8.528802 | 0.554548 | 9.687705 | 3.482576 | 0         | 0        |
|     | Rank | 3        | 7        | 9        | 5        | 10       | 4        | 8        | 6        | 2         | 1        |
| F5  | Avg  | 28.25658 | 172.1007 | 147484.5 | 54.16325 | 3517.504 | 170.2001 | 92095.15 | 109.1053 | 3.237979  | 1.622074 |
|     | Std  | 0.406265 | 392.8285 | 142560.8 | 68.85537 | 16368.69 | 200.7955 | 374414.4 | 164.4389 | 8.395855  | 6.171846 |
|     | Rank | 3        | 7        | 10       | 4        | 8        | 6        | 9        | 5        | 2         | 1        |
| F6  | Avg  | 2.80945  | 1.06E-05 | 1120.928 | 0        | 1656.709 | 0.31324  | 0.363056 | 9.27E-08 | 1.76E-07  | 6.39E-08 |
|     | Std  | 0.233264 | 9.56E-06 | 509.9442 | 0        | 3767.988 | 0.070063 | 0.128503 | 2.79E-09 | 2.29E-07  | 6.87E-08 |
|     | Rank | 8        | 5        | 9        | 1        | 10       | 6        | 7        | 3        | 4         | 2        |
| F7  | Avg  | 4.86E-05 | 0.09513  | 0.016716 | 0.057969 | 1.63063  | 0.018222 | 0.031431 | 0.096089 | 7.79E-05  | 3.35E-05 |
|     | Std  | 4.52E-05 | 0.035386 | 0.013431 | 0.024491 | 2.99583  | 0.00738  | 0.028207 | 0.032296 | 9.81E-05  | 4.14E-05 |
|     | Rank | 2        | 8        | 4        | 7        | 10       | 5        | 6        | 9        | 3         | 1        |
| F8  | Avg  | -3202.86 | -2413.53 | -5754.17 | -1587.84 | -3140.14 | -3025.94 | -2324.05 | -2883.38 | -12569.5  | -12569.5 |
|     | Std  | 240.9948 | 496.7874 | 591.5846 | 300.7621 | 347.8142 | 290.339  | 202.7524 | 311.6512 | 0.000145  | 4.66E-05 |
|     | Rank | 4        | 8        | 3        | 10       | 5        | 6        | 9        | 7        | 2         | 1        |
| F9  | Avg  | 0        | 80.45882 | 162.1103 | 25.37144 | 161.282  | 106.9613 | 17.15178 | 57.17687 | 0         | 0        |
|     | Std  | 0        | 25.80236 | 38.31689 | 5.746865 | 43.39013 | 32.2269  | 22.96072 | 22.52042 | 0         | 0        |
|     | Rank | 1        | 5        | 8        | 3        | 7        | 6        | 2        | 4        | 1         | 1        |
| F10 | Avg  | 8.88E-16 | 1.973304 | 9.232492 | 8.40E-09 | 14.00294 | 1.283419 | 13.96645 | 2.165787 | 8.88E-16  | 8.88E-16 |
|     | Std  | 2.01E-31 | 0.611315 | 1.44599  | 2.20E-09 | 7.950571 | 0.728616 | 8.91743  | 0.895879 | 0         | 0        |
|     | Rank | 2        | 5        | 7        | 3        | 9        | 4        | 8        | 6        | 1         | 1        |
| F11 | Avg  | 0.101451 | 0.01314  | 12.75532 | 9.167486 | 15.07443 | 0.550285 | 0.29508  | 0.006158 | 0         | 0        |
|     | Std  | 0.079528 | 0.014247 | 6.049923 | 5.613008 | 53.42284 | 0.110778 | 0.275885 | 0.006458 | 0         | 0        |
|     | Rank | 4        | 3        | 8        | 7        | 9        | 6        | 5        | 2        | 1         | 1        |
| F12 | Avg  | 0.390268 | 11.79996 | 337.437  | 0.147818 | 0.842388 | 1.671162 | 100.5139 | 6.031526 | 4.07E-08  | 1.27E-08 |
|     | Std  | 0.05154  | 3.893738 | 1274.537 | 0.190267 | 1.086202 | 1.03551  | 528.101  | 3.243773 | 6.33E-08  | 1.43E-08 |
|     | Rank | 4        | 8        | 10       | 3        | 5        | 6        | 9        | 7        | 2         | 1        |
| F13 | Avg  | 2.775434 | 3.566452 | 60001.51 | 0.039062 | 0.686478 | 0.057376 | 107.5754 | 2.688221 | 0.004358  | 0.001465 |
|     | Std  | 0.105293 | 11.30312 | 157203.5 | 0.163974 | 1.111076 | 0.033767 | 475.3857 | 8.903576 | 0.010949  | 0.003799 |
|     | Rank | 7        | 8        | 10       | 3        | 5        | 4        | 9        | 6        | 2         | 1        |
| F14 | Avg  | 9.305394 | 1.428617 | 0.998004 | 3.920338 | 1.691591 | 0.998004 | 1.655152 | 1.031138 | 0.998004  | 0.998004 |
|     | Std  | 3.82843  | 0.564551 | 6.39E-08 | 2.721899 | 1.165462 | 6.78E-16 | 1.87623  | 0.181484 | 4.12E-17  | 0        |
|     | Rank | 10       | 6        | 4        | 9        | 8        | 3        | 7        | 5        | 2         | 1        |
| F15 | Avg  | 0.010175 | 0.004013 | 0.00371  | 0.002483 | 0.001622 | 0.003847 | 0.000862 | 0.001416 | 0.000491  | 0.000407 |
|     | Std  | 0.022732 | 0.007439 | 0.005836 | 0.00146  | 0.003566 | 0.007011 | 0.000335 | 0.003584 | 0.000373  | 0.000278 |
|     | Rank | 10       | 9        | 7        | 6        | 5        | 8        | 3        | 4        | 2         | 1        |
| F16 | Avg  | -1.03163 | -1.03163 | -1.03163 | -1.03163 | -1.03163 | -1.03163 | -1.0316  | -1.03163 | -1.03163  | -1.03163 |
|     | Std  | 9.32E-08 | 0        | 3.66E-06 | 0        | 0        | 1.04E-07 | 3.03E-05 | 0        | 6.25E-16  | 0        |
|     | Rank | 3        | 1        | 5        | 1        | 1        | 4        | 6        | 1        | 2         | 1        |
| F17 | Avg  | 0.405421 | 0.397887 | 0.397895 | 0.397887 | 0.397887 | 0.397888 | 0.399022 | 0.397887 | 0.397887  | 0.397887 |
|     | Std  | 0.005929 | 1.69E-16 | 3.56E-05 | 1.69E-16 | 1.69E-16 | 2.06E-07 | 0.001427 | 1.69E-16 | 0         | 0        |
|     | Rank | 6        | 2        | 4        | 2        | 2        | 3        | 5        | 2        | 1         | 1        |
| F18 | Avg  | 10.0868  | 3        | 3.000007 | 3        | 3        | 3.000001 | 3.000021 | 3        | 3         | 3        |
|     | Std  | 11.95653 | 0        | 2.24E-05 | 0        | 0        | 7.04E-07 | 3.49E-05 | 0        | 1.04E-15  | 0        |
|     | Rank | 6        | 1        | 5        | 1        | 1        | 3        | 4        | 1        | 2         | 1        |
| F19 | Avg | -3.85342 | -3.86278 | -3.8627 | -3.86278 | -3.86252 | -3.86278 | -3.85498 | -3.86278 | -3.86278 | -3.86278 |
| | Std | 0.002905 | 1.36E-15 | 0.000138 | 1.36E-15 | 0.001439 | 3.97E-07 | 0.001994 | 1.36E-15 | 2.65E-15 | 1.36E-15 |
| | Rank | 7 | 1 | 3 | 1 | 5 | 4 | 6 | 1 | 2 | 1 |
| F20 | Avg | -3.09355 | -3.27839 | -3.24973 | -3.22521 | -3.2341 | -3.25042 | -2.91013 | -3.22916 | -3.27444 | -3.29822 |
| | Std | 0.085726 | 0.058284 | 0.077816 | 0.09226 | 0.060746 | 0.059439 | 0.377887 | 0.052154 | 0.05837 | 0.049241 |
| | Rank | 9 | 2 | 5 | 8 | 6 | 4 | 10 | 7 | 3 | 1 |
| F21 | Avg | -3.74402 | -6.53424 | -7.35707 | -5.59336 | -6.14456 | -7.46314 | -2.82024 | -8.13991 | -10.1532 | -10.1532 |
| | Std | 0.788914 | 2.932695 | 2.673835 | 3.575389 | 3.448257 | 3.023758 | 1.808205 | 2.980425 | 6.08E-15 | 6.04E-15 |
| | Rank | 9 | 6 | 5 | 8 | 7 | 4 | 10 | 3 | 2 | 1 |
| F22 | Avg | -4.0787 | -7.66281 | -8.67404 | -10.27 | -8.42414 | -9.16675 | -3.43843 | -10.2271 | -10.4029 | -10.4029 |
| | Std | 1.044784 | 3.22496 | 2.679209 | 0.728055 | 3.141982 | 2.278944 | 2.01902 | 0.962918 | 5.71E-16 | 4.33E-16 |
| | Rank | 9 | 8 | 6 | 3 | 7 | 5 | 10 | 4 | 2 | 1 |
| F23 | Avg | -3.96481 | -7.02753 | -8.36917 | -10.4662 | -7.89781 | -9.40438 | -4.84152 | -8.96732 | -10.5364 | -10.5364 |
| | Std | 1.469116 | 3.438882 | 2.905126 | 0.384508 | 3.584177 | 2.618164 | 2.075541 | 3.192975 | 1.28E-15 | 1.22E-15 |
| | Rank | 10 | 7 | 6 | 3 | 8 | 4 | 9 | 5 | 2 | 1 |
| Sum of Rank | | 122 | 126 | 151 | 99 | 155 | 109 | 160 | 102 | 44 | 24 |
| Average Rank | | 12.2 | 12.6 | 15.1 | 9.9 | 15.5 | 10.9 | 16 | 10.2 | 4.4 | 2.4 |
| Final Rank | | 6 | 7 | 8 | 3 | 9 | 5 | 10 | 4 | 2 | 1 |

It can be seen from Table 3 that the AGTO algorithm outperforms the other state of art algorithms.

### 5.1.3. Statistical analysis of AGTO algorithm

The comparison of the algorithms has been over 30 independent runs in terms of Avg and Std. But the individual run is not analysed during this comparison. So there can be a possibility of getting superiority by chance. Thus it is necessary to compare each run to validate its significance. For this purpose, the Wilcoxon rank sum test is carried out at a 95% confidence level. The P values have been calculated as shown in Table 4. If the obtained P-value is less than 0.05 then it suggests the null hypothesis which means the best algorithm gives a higher value for the objective function that not happened by the accident. The best algorithm based on the least value of Std. is selected and compared to other algorithms separately for statistical analysis. The best algorithm is labelled as N/A as it cannot be compared to itself.

**Table 4**
Results of the Wilcoxon rank-sum test

| | AOA | ALO | DA | GWO | MFO | MVO | SCA | SSA | GTO | AGTO |
|---|---|---|---|---|---|---|---|---|---|---|
| F1 | $4.50 \times 10^{-11}$ | $4.50 \times 10^{-11}$ | $3.02 \times 10^{-11}$ | $3.02 \times 10^{-11}$ | $2.98 \times 10^{-11}$ | $3.02 \times 10^{-11}$ | $3.02 \times 10^{-11}$ | $3.02 \times 10^{-11}$ | NaN | NaN |
| F2 | NaN | $1.21 \times 10^{-12}$ | $1.21 \times 10^{-12}$ | $1.21 \times 10^{-12}$ | $1.21 \times 10^{-12}$ | $1.21 \times 10^{-12}$ | $1.21 \times 10^{-12}$ | $1.21 \times 10^{-12}$ | NaN | NaN |
| F3 | $3.34 \times 10^{-11}$ | $3.02 \times 10^{-11}$ | $3.02 \times 10^{-11}$ | $3.02 \times 10^{-11}$ | $3.02 \times 10^{-11}$ | $3.02 \times 10^{-11}$ | $3.02 \times 10^{-11}$ | $3.02 \times 10^{-11}$ | NaN | NaN |
| F4 | $3.34 \times 10^{-11}$ | $3.02 \times 10^{-11}$ | $3.02 \times 10^{-11}$ | $3.02 \times 10^{-11}$ | $3.02 \times 10^{-11}$ | $3.02 \times 10^{-11}$ | $3.02 \times 10^{-11}$ | $3.02 \times 10^{-11}$ | 0.333711 | NaN |
| F5 | $3.02 \times 10^{-11}$ | $3.02 \times 10^{-11}$ | $3.02 \times 10^{-11}$ | $3.02 \times 10^{-11}$ | $3.02 \times 10^{-11}$ | $3.02 \times 10^{-11}$ | $3.02 \times 10^{-11}$ | $3.02 \times 10^{-11}$ | $3.02 \times 10^{-11}$ | NaN |
| F6 | $1.21 \times 10^{-12}$ | $1.21 \times 10^{-12}$ | $1.21 \times 10^{-12}$ | NaN | $1.20 \times 10^{-12}$ | $1.21 \times 10^{-12}$ | $1.21 \times 10^{-12}$ | $1.21 \times 10^{-12}$ | $1.21 \times 10^{-12}$ | $1.21 \times 10^{-12}$ |
| F7 | 0.032651 | $3.02 \times 10^{-11}$ | $3.02 \times 10^{-11}$ | $3.02 \times 10^{-11}$ | $3.02 \times 10^{-11}$ | $3.02 \times 10^{-11}$ | $3.02 \times 10^{-11}$ | $3.02 \times 10^{-11}$ | 0.032651 | NaN |
| F8 | $3.02 \times 10^{-11}$ | $2.99 \times 10^{-11}$ | $3.02 \times 10^{-11}$ | $3.02 \times 10^{-11}$ | $2.99 \times 10^{-11}$ | $3.02 \times 10^{-11}$ | $3.02 \times 10^{-11}$ | $3.02 \times 10^{-11}$ | 0.032651 | NaN |
| F9 | NaN | $1.21 \times 10^{-12}$ | $1.21 \times 10^{-12}$ | $1.20 \times 10^{-12}$ | $1.21 \times 10^{-12}$ | $1.21 \times 10^{-12}$ | $1.21 \times 10^{-12}$ | $1.21 \times 10^{-12}$ | NaN | NaN |
| F10 | NaN | $1.21 \times 10^{-12}$ | $1.21 \times 10^{-12}$ | $1.21 \times 10^{-12}$ | $1.21 \times 10^{-12}$ | $1.21 \times 10^{-12}$ | $1.21 \times 10^{-12}$ | $1.20 \times 10^{-12}$ | NaN | NaN |
| F11 | $1.21 \times 10^{-12}$ | $1.21 \times 10^{-12}$ | $1.21 \times 10^{-12}$ | $1.21 \times 10^{-12}$ | $1.21 \times 10^{-12}$ | $1.21 \times 10^{-12}$ | $1.21 \times 10^{-12}$ | $1.21 \times 10^{-12}$ | NaN | NaN |
| F12 | $3.02 \times 10^{-11}$ | $3.02 \times 10^{-11}$ | $3.02 \times 10^{-11}$ | 0.379036 | $3.02 \times 10^{-11}$ | $3.02 \times 10^{-11}$ | $3.02 \times 10^{-11}$ | $3.02 \times 10^{-11}$ | $5.57 \times 10^{-10}$ | NaN |
| F13 | $3.02 \times 10^{-11}$ | $4.98 \times 10^{-11}$ | $3.02 \times 10^{-11}$ | 0.185253 | $3.02 \times 10^{-11}$ | $3.02 \times 10^{-11}$ | $3.02 \times 10^{-11}$ | 0.662682 | 0.662572 | **NaN** |
| F14 | $8.41 \times 10^{-13}$ | 0.000132 | 0.333711 | $4.57 \times 10^{-12}$ | 0.00031 | $8.41 \times 10^{-12}$ | $4.50 \times 10^{-12}$ | 0.333711 | 0.021561 | **NaN** |
| F15 | $7.20 \times 10^{-05}$ | $6.52 \times 10^{-09}$ | $3.69 \times 10^{-11}$ | $3.02 \times 10^{-11}$ | $7.32 \times 10^{-11}$ | $1.09 \times 10^{-10}$ | $1.33 \times 10^{-10}$ | $6.07 \times 10^{-11}$ | $2.67 \times 10^{-09}$ | **NaN** |
| F16 | $2.39 \times 10^{-12}$ | **NaN** | $1.24 \times 10^{-07}$ | **NaN** | **NaN** | $3.71 \times 10^{-13}$ | $1.21 \times 10^{-12}$ | **NaN** | $1.21 \times 10^{-12}$ | **NaN** |
| F17 | $1.21 \times 10^{-12}$ | $1.21 \times 10^{-12}$ | $1.27 \times 10^{-05}$ | $1.21 \times 10^{-12}$ | $1.21 \times 10^{-12}$ | $4.48 \times 10^{-12}$ | $1.21 \times 10^{-12}$ | $1.21 \times 10^{-12}$ | **NaN** | **NaN** |
| F18 | 0.002787 | **NaN** | $1.27 \times 10^{-05}$ | **NaN** | **NaN** | $5.70 \times 10^{-11}$ | $1.21 \times 10^{-12}$ | **NaN** | $1.21 \times 10^{-12}$ | **NaN** |
| F19 | $1.21 \times 10^{-12}$ | **NaN** | $4.57 \times 10^{-12}$ | **NaN** | 0.333711 | $1.67 \times 10^{-08}$ | $1.21 \times 10^{-12}$ | **NaN** | $1.21 \times 10^{-12}$ | **NaN** |
| F20 | $1.21 \times 10^{-12}$ | 0.000313 | $4.57 \times 10^{-12}$ | $1.21 \times 10^{-12}$ | $2.92 \times 10^{-08}$ | $1.21 \times 10^{-12}$ | $1.21 \times 10^{-12}$ | $5.85 \times 10^{-09}$ | $4.79 \times 10^{-08}$ | **NaN** |
| F21 | $3.02 \times 10^{-11}$ | 0.075837 | $2.87 \times 10^{-06}$ | 0.074498 | 0.183369 | 0.147367 | $3.02 \times 10^{-11}$ | 0.024255 | 0.001489 | **NaN** |
| F22 | $3.02 \times 10^{-11}$ | 0.372965 | 0.001679 | $4.56 \times 10^{-11}$ | 0.006661 | 0.003938 | $3.02 \times 10^{-11}$ | $4.56 \times 10^{-11}$ | 0.003562 | **NaN** |
| F23 | $3.02 \times 10^{-11}$ | 0.660515 | $8.19 \times 10^{-07}$ | $4.56 \times 10^{-11}$ | 0.072549 | 0.000168 | $3.02 \times 10^{-11}$ | $3.81 \times 10^{-05}$ | 0.06675 | **NaN** |

## 5.2. Need to optimize the hyperparameters of CNN.

The accuracy of the CNN mainly depends on the optimal selection of its hyperparameters. The improper selection of these hyperparameters of the built CNN model leads to suboptimal results because it fails to minimize the loss function. Some of the hyperparameters of the CNN are set in advance to reduce the search space and computational time of AGTO. The size of the input image is 656x875. The input image is down-sampled to 2X2 by the max pooling layer. Contrary to the AlexNet model, the constructed CNN does not employ a dropout approach between the fully connected layers because there aren't many deeper layers.

There are other hyperparameters also which influence the performance of the CNN. The optimal selection of these hyperparameters becomes very important as these parameters maximize the performance of the model and, at the same time, minimize the loss function to produce good results with fewer errors. The hyperparameters which have been considered in the present research are the number of neurons in the hidden layer, learning rate, batch size, epoch, and activation function. As the learning rate controls the step size, which is required to achieve the minimum loss function. With a higher learning rate, the model learns more quickly, but it may not achieve the minimum loss function resulting in merely reaching its surroundings. The minimum loss functions can be easily achieved with a lower learning rate. Thus it becomes necessary to choose the learning rate wisely to make the model capable enough to learn at a faster rate with a minimum loss function. If the training dataset is very large, the time required to build the CNN model will be higher. Thus, assigning the batch size to the training datasets prevents the built model from receiving the data at once, which speeds up the learning process. With reduced batch size, the learning process becomes faster, but the accuracy of the model decreases while checking with validation data. Thus the optimal batch should be chosen to balance between learning rate and accuracy. Epochs represent the number of times a complete dataset is run in the CNN model. If fewer numbers of epochs have been selected, then it will result in underfitting of the data, while a higher number of epochs results in overfitting of the data. Therefore, to get the best outcome of the built CNN model, it is necessary to select the optimum number of epochs.

The range of hyperparameters which have been considered for optimization is tabulated in Table 5.

**Table 5**
Range of Hyperparameters in AGTO

| S.No. | Hyperparameters | Range |
|---|---|---|
| 1 | No. neurons in the hidden layers | [10,100] |
| 2 | Learning rate | [0.01, 1] |
| 3 | Batch size | [200, 1000] |
| 4 | Epoch | [2, 100] |
| 5 | Activation function | [0,9] |

The range of activation function is [0,9], where 0 to 9 represents different activation functions such as relu, sigmoid, softplus, softsign, tanh, selu, elu, exponential, leakyrelu and prelu.

### 5.3. Analyzing the diagnosis results

The basic GTO has been modified by incorporating the opposition-based learning concept and quantum gate rotation mutation strategy. Further, the amended GTO algorithm has been used to construct the CNN model by optimizing its hyperparameters. The Gaussian process has been utilized to update the posterior distribution of the objective function, whereas the expected improvement is selected as the acquisition function for effective sampling. The classification accuracy of the built CNN model, along with its loss function, is shown in Fig. 11.

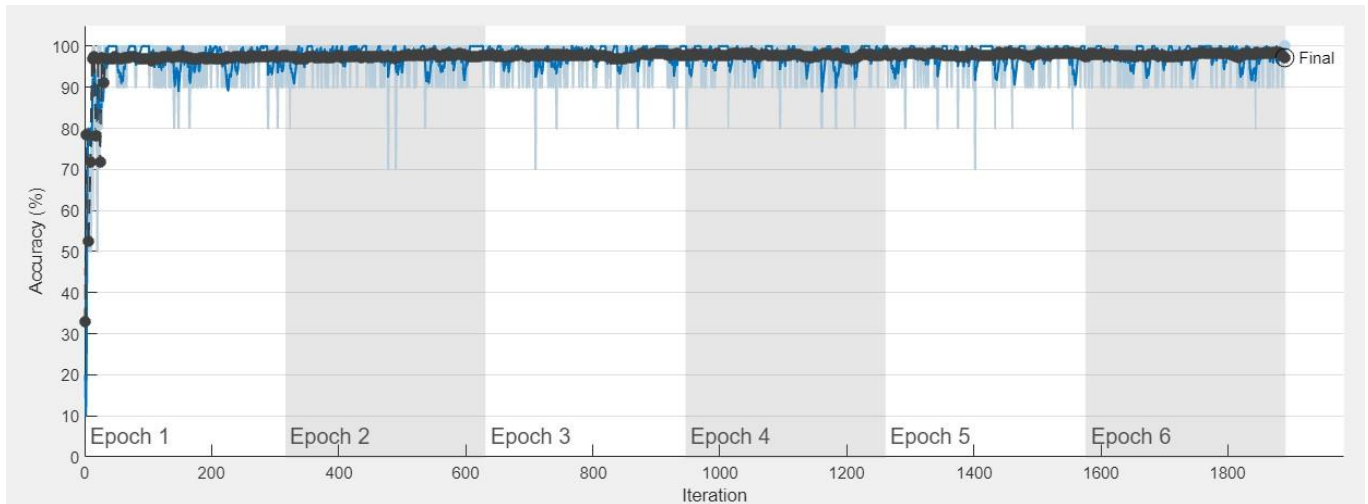

(a)

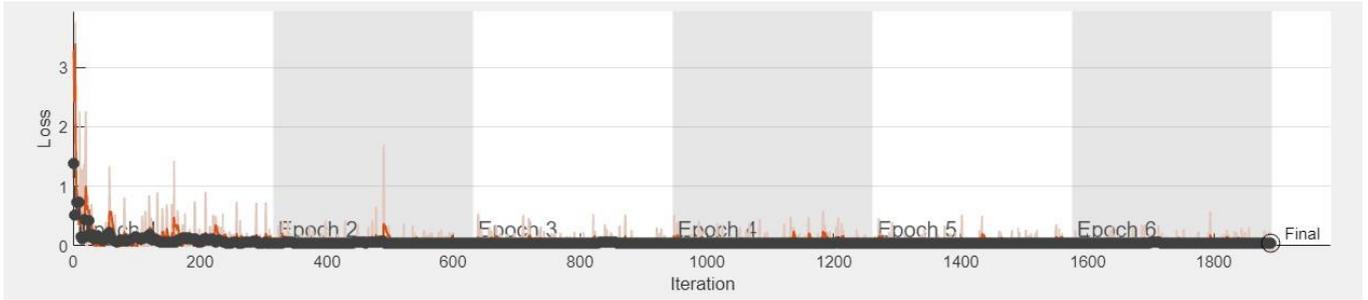

**(b)**

**Fig. 11** Performance of AGTO-CNN model (a) accuracy (b) loss

The performance of the built CNN model has been verified by repeated trials. During repeated trials, very small fluctuations have been observed, which means the built AGTO-CNN converges well and presents good stability. The average test accuracy of the repeated trials of the proposed methodology has been compared with other models of CNN. The result of the comparison is shown in Fig. 12 and tabulated in Table 6. The standard deviation has also been computed and presented in Table 6. It can be seen that the AGTO-CNN have the lowest value of standard deviation when compared to other models. Apart from other models of CNN, the proposed AGTO-CNN has been compared with other classifiers such as ANN, ANFIS, SVM, ELM and Neuro Fuzzy as shown in Fig. 13. The proposed scheme has been checked with different optimization algorithms viz., salp swarm algorithm (SSA), arithmetic optimization algorithm (AOA), sine cosine algorithm (SCA), gravitational search algorithm (GSA), and gorilla troop optimization (GTO) as shown in Fig. 14. It is observed from the results of the comparison that AGTO-CNN not only gives a higher degree of accuracy but also gives the least standard deviation.

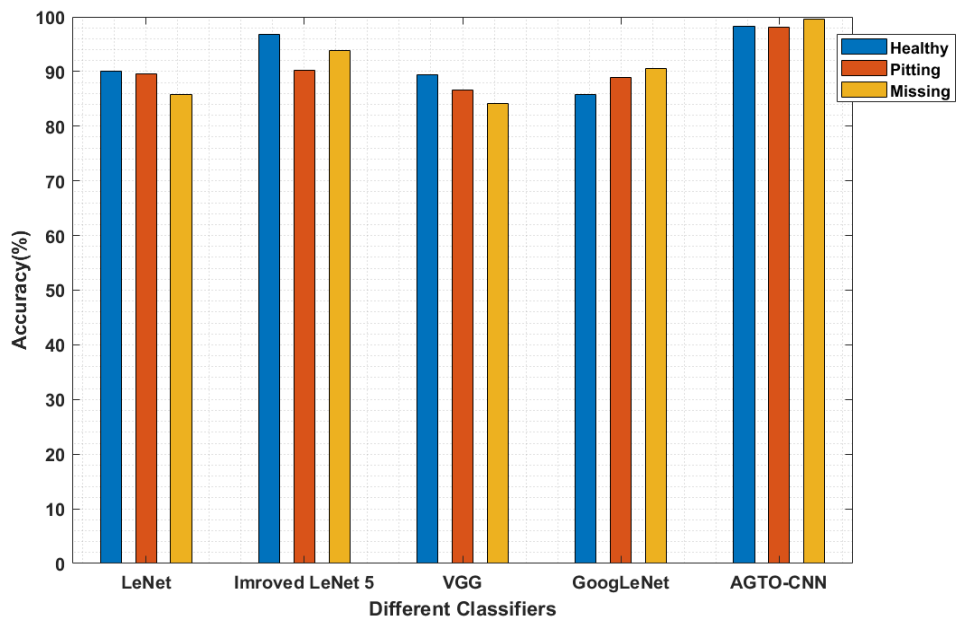

**Fig. 12** Comparison of the AGTO-CNN model with the other existing models of CNN

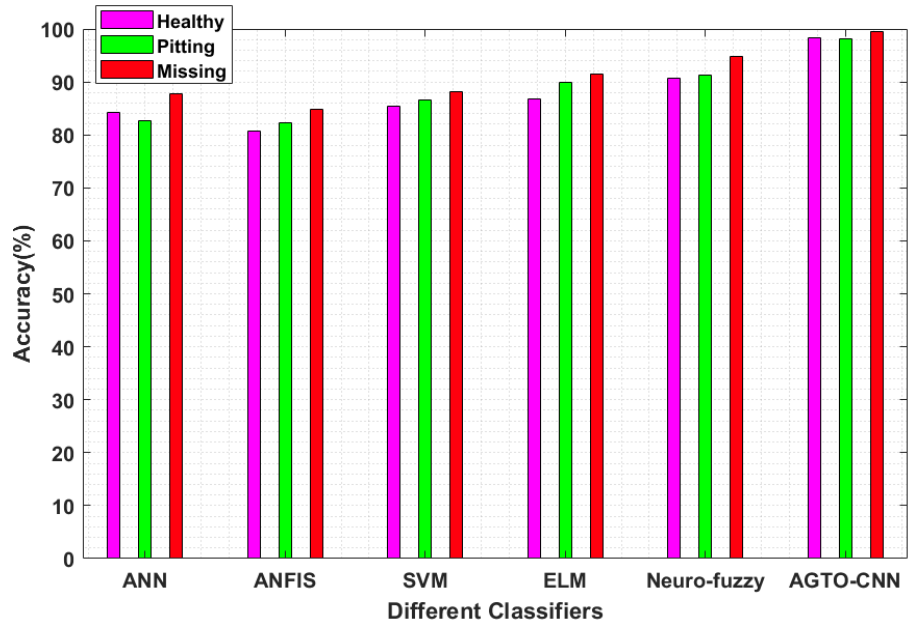

**Fig. 13** Comparison of the AGTO-CNN model with other classifiers

**Table 6**
Comparison of accuracy and Std with different models

| Model | Test Accuracy (%) | Standard deviation (Std) |
| --- | --- | --- |
| LeNet | 88.58 | 0.9658 |
| Improved LeNet 5 | 95.63 | 0.3521 |
| VGG | 90.56 | 0.4062 |
| GoogLeNet | 92.38 | 0.3714 |
| Proposed AGTO-CNN | 98.95 | 0.2145 |

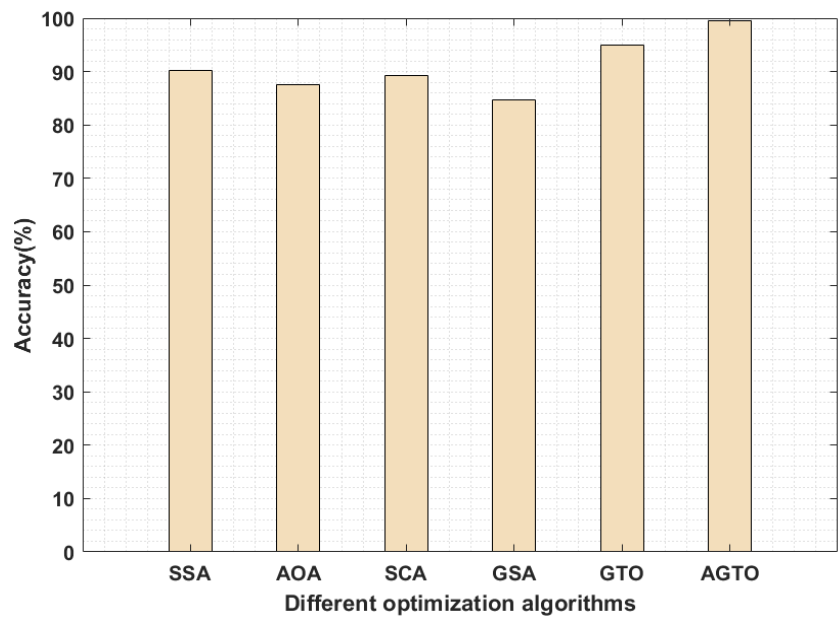

**Fig. 14** Comparison of the AGTO with other optimization algorithms in terms of recognition accuracy

The confusion matrix has also been generated for one trial to examine the recognition accuracy of each health state as shown in Fig. 15. The recognition accuracy for each health condition is tabulated in Table 7. It can easily be seen from Table 7 that the built model correctly identified each health state of the worm wheel.

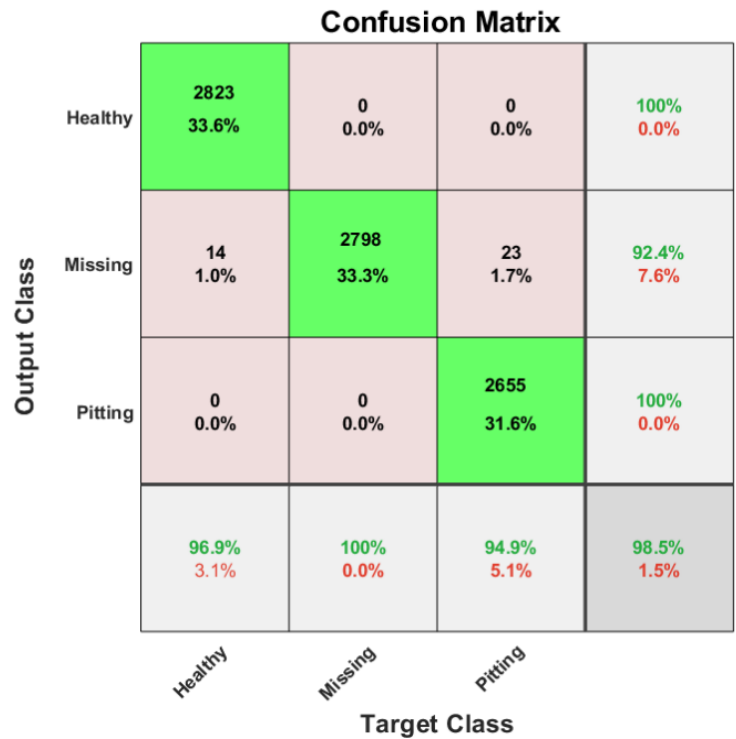

**Fig. 15** Confusion matrix

**Table 7**
Recognition of different worm gearbox defects by different methods

| Different Health States | LeNet | Improved LeNet | VGG | GoogLeNet | AGTO-CNN |
|---|---|---|---|---|---|
| **Healthy** | 90.12 | 96.78 | 89.45 | 85.72 | 98.32 |
| **Pitting** | 89.62 | 90.23 | 86.62 | 88.85 | 98.12 |
| **Missing** | 85.72 | 93.89 | 84.15 | 90.56 | 99.59 |

It is also necessary to analyse the computational complexity of the proposed AGTO-CNN to validate its efficacy. Computational complexity consists of both time complexity and spatial complexity. The time complexity generally determines the training and prediction time of the built model of AGTO-CNN. At the same time, spatial complexity is used to tell the number of parameters of the built model. The more parameters a model has, the more data it may need to train because of the limitation of the "curse of dimensionality." If $L$ represents the side length of the feature graph obtained from the convolution kernel, $K$ is the length of each convolution kernel, $Z_{in}$ and $Z_{out}$ are input and output channels respectively and $W$ is the input size. Then the computational complexity of AGTO-CNN is tabulated in Table 8.

**Table 8**
Computational complexity of AGTO-CNN

| Layer | $L$ | $K$ | $Z_{in}$ | $Z_{out}$ | Time complexity | Spatial complexity |
|---|---|---|---|---|---|---|
| Conv1 | 70 | 6 | 4 | 9 | 2430000 | 33075 |
| Conv2 | 25 | 5 | 9 | 15 | 2028000 | 16875 |
| FC1 | 1 | 12 | 15 | 380 | 1091258 | 1091258 |
| FC2 | 1 | 1 | 380 | 150 | 57000 | 57000 |
| Output layer | 1 | 1 | 150 | 5 | 750 | 750 |
| Total |  |  |  |  | 5607008 | 1198958 |

It has been confirmed from the above analysis that the AGTO-CNN is a robust model which gives a higher degree of accuracy and presents good stability. The proposed methodology can also be used to detect faults or anomalies as early as possible in other applications. This information helps the maintenance personnel to resolve the defects before further damage. In this way, the operational efficiency of the machine can be improved by reducing energy consumption and by increasing the reliability of the systems simultaneously.

## 6. Conclusions

In this research, an amended gorilla troop optimization (AGTO) has been proposed by integrating the opposition-based learning concept and quantum gate rotation mutation strategy to address the problems of stucking the solution in local optima and slow convergence. The proposed AGTO has been tested against the different optimization algorithms on twenty-three basic benchmark functions in terms of Avg and Std. AGTO gave improved results for different benchmark functions when compared to other optimization algorithms. The statistical analysis of the proposed optimization algorithm has been carried out through the Wilcoxon test, which indicated that the proposed algorithm is significant at most of the benchmark functions.

The developed AGTO algorithm is utilized to make the CNN adaptive by optimizing its key HPs. The developed AGTO-CNN model is further used to detect the different defects of the worm gearbox. First, the vibration and acoustic signals acquired from the worm gearbox test rig are transformed into time-frequency images through the Morlet wavelet function. The built AGTO-CNN model learns the prominent features from the time-frequency images and classifies the different defects of the worm gearbox. The developed AGTO-CNN model has fast convergence, stability and strong robustness as it replaced the traditional method of manual adjustment of key HPs of CNN by automatic optimization. The diagnostic accuracy of AGTO-CNN is 98.95%, which is more than that of traditional methods. A confusion matrix has also been constructed for the different health conditions. The results obtained revealed that AGTO-CNN is able enough to learn the hidden characteristics embedded in the time-frequency images.

In future, the proposed methodology can be extended to diagnose the other mechanical components such as bearings, shafts, rotors etc.

**Acknowledgements**
- The work of Radoslaw Zimroz is supported by the National Center of Science, Poland under Sheng2 project No. UMO-2021/40/Q/ST8/00024 "NonGauMech - New methods of processing non-stationary signals (identification, segmentation, extraction, modeling) with non-Gaussian characteristics for the purpose of monitoring complex mechanical structures".
- The authors gratefully acknowledge the Precision Metrology Laboratory, Department of Mechanical Engineering of Sant Longowal Institute of Engineering and Technology Longowal, India for carrying out the experimentation.

**References**

[1] Tan, H. *et al.* Sensible multiscale symbol dynamic entropy for fault diagnosis of bearing. *Int. J. Mech. Sci.* **256**, 108509 (2023).

[2] Wu, Z., Jiang, H., Zhu, H. & Wang, X. A knowledge dynamic matching unit-guided multi-source domain adaptation network with attention mechanism for rolling bearing fault diagnosis. *Mech. Syst. Signal Process.* **189**, 110098 (2023).

[3] Wen, H., Guo, W. & Li, X. A novel deep clustering network using multi-representation autoencoder and adversarial learning for large cross-domain fault diagnosis of rolling bearings. *Expert Syst. Appl.* **225**, 120066 (2023).

[4] Kumar, A. & Kumar, R. Role of Signal Processing, Modeling and Decision Making in the Diagnosis of Rolling Element Bearing Defect: A Review. *J. Nondestruct. Eval.* **38**, 1–29 (2019).

[5] Tahan, M., Tsoutsanis, E., Muhammad, M. & Abdul Karim, Z. A. Performance-based health monitoring, diagnostics and prognostics for condition-based maintenance of gas turbines: A review. *Appl. Energy* **198**, 122–144 (2017).

[6] Valente de Oliveira, J. *et al.* A review on data-driven fault severity assessment in rolling bearings. *Mech. Syst. Signal Process.* **99**, 169–196 (2017).

[7] Lei, Y. *et al.* Applications of machine learning to machine fault diagnosis: A review and roadmap. *Mech. Syst. Signal Process.* **138**, 106587 (2020).

[8] Tang, X. *et al.* Intelligent fault diagnosis of helical gearboxes with compressive sensing based non-contact measurements. *ISA Trans.* **133**, 559–574 (2023).

[9] Meng, L. *et al.* Intelligent fault diagnosis of gearbox based on differential continuous

wavelet transform-parallel multi-block fusion residual network. *Measurement* **206**, 112318 (2023).

[10] Kumar, A., Gandhi, C. P., Zhou, Y., Kumar, R. & Xiang, J. Latest developments in gear defect diagnosis and prognosis: A review. *Measurement* **158**, 107735 (2020).

[11] Kumar, S. & Kumar, R. Diagnosis of an incipient defect in a worm gearbox using minimum entropy deconvolution and local cepstrum. *Meas. Sci. Technol.* **32**, 054002 (2021).

[12] Kumar, S. & Kumar, R. L-Moments Ratio-Based Condition Indicators for Diagnosis of Fault in a Worm Gearbox. *J. Vib. Eng. Technol.* (2023). doi:10.1007/s42417-022-00807-2

[13] Koutsoupakis, J., Seventekidis, P. & Giagopoulos, D. Machine learning based condition monitoring for gear transmission systems using data generated by optimal multibody dynamics models. *Mech. Syst. Signal Process.* **190**, 110130 (2023).

[14] Vashishtha, G. & Kumar, R. An effective health indicator for the Pelton wheel using a Levy flight mutated genetic algorithm. *Meas. Sci. Technol.* **32**, (2021).

[15] Lou, Y., Kumar, A. & Xiang, J. Machinery Fault Diagnosis Based on Domain Adaptation to Bridge the Gap Between Simulation and Measured Signals. *IEEE Trans. Instrum. Meas.* **71**, 1–9 (2022).

[16] Shi, Q. & Zhang, H. Fault Diagnosis of an Autonomous Vehicle with an Improved SVM Algorithm Subject to Unbalanced Datasets. *IEEE Trans. Ind. Electron.* **68**, 6248–6256 (2021).

[17] Gao, Y., Liu, X. & Xiang, J. Fault Detection in Gears Using Fault Samples Enlarged by a Combination of Numerical Simulation and a Generative Adversarial Network. *IEEE/ASME Trans. Mechatronics* **27**, 3798–3805 (2022).

[18] Vashishtha, G., Chauhan, S., Kumar, A. & Kumar, R. An ameliorated African vulture optimization algorithm to diagnose the rolling bearing defects. *Meas. Sci. Technol.* **33**, (2022).

[19] Gao, Y., Liu, X. & Xiang, J. FEM Simulation-Based Generative Adversarial Networks to Detect Bearing Faults. *IEEE Trans. Ind. Informatics* **16**, 4961–4971 (2020).

[20] Tang, S., Zhu, Y. & Yuan, S. Intelligent fault diagnosis of hydraulic piston pump based on deep learning and Bayesian optimization. *ISA Trans.* (2022). doi:10.1016/j.isatra.2022.01.013

[21] Kumar, A. *et al.* Novel Convolutional Neural Network ( NCNN ) for the Diagnosis of Bearing Defects in Rotary Machinery. *IEEE Trans. Instrum. Meas.* **70**, (2021).

[22] Vashishtha, G. & Kumar, R. Unsupervised Learning Model of Sparse Filtering Enhanced Using Wasserstein Distance for Intelligent Fault Diagnosis. *J. Vib. Eng. Technol.* (2022). doi:10.1007/s42417-022-00725-3

[23] Wu, J., Tang, T., Chen, M., Wang, Y. & Wang, K. A study on adaptation lightweight architecture based deep learning models for bearing fault diagnosis under varying working conditions. *Expert Syst. Appl.* **160**, 113710 (2020).

[24] Zhao, B., Zhang, X., Li, H. & Yang, Z. Intelligent fault diagnosis of rolling bearings based on normalized CNN considering data imbalance and variable working conditions. *Knowledge-Based Syst.* **199**, 105971 (2020).

[25] Zhu, Z., Peng, G., Chen, Y. & Gao, H. A convolutional neural network based on a capsule network with strong generalization for bearing fault diagnosis. *Neurocomputing* **323**, 62–75 (2019).

[26] Gai, J., Shen, J., Hu, Y. & Wang, H. An integrated method based on hybrid grey wolf optimizer improved variational mode decomposition and deep neural network for fault diagnosis of rolling bearing. *Measurement* **162**, 107901 (2020).

[27] Surucu, O., Gadsden, S. A. & Yawney, J. Condition Monitoring using Machine Learning: A Review of Theory, Applications, and Recent Advances. *Expert Syst. Appl.* **221**, 119738 (2023).

[28] Abd Elaziz, M., Abualigah, L., Issa, M. & Abd El-Latif, A. A. Optimal parameters extracting of fuel cell based on Gorilla Troops Optimizer. *Fuel* **332**, 126162 (2023).

[29] Abdel-Basset, M., El-Shahat, D., Sallam, K. M. & Munasinghe, K. Parameter extraction of photovoltaic models using a memory-based improved gorilla troops optimizer. *Energy Convers. Manag.* **252**, 115134 (2022).

[30] Piri, J. *et al.* Feature Selection Using Artificial Gorilla Troop Optimization for Biomedical Data: A Case Analysis with COVID-19 Data. *Mathematics* **10**, 1–31 (2022).

[31] Ginidi, A. *et al.* Gorilla troops optimizer for electrically based single and double-diode models of solar photovoltaic systems. *Sustainability* **13**, (2021).

[32] Vashishtha, G. & Kumar, R. An amended grey wolf optimization with mutation strategy to diagnose bucket defects in Pelton wheel. *Measurement* **187**, 110272 (2022).

[33] Jia, L., Chow, T. W. S. & Yuan, Y. GTFE-Net: A Gramian Time Frequency Enhancement CNN for bearing fault diagnosis. *Eng. Appl. Artif. Intell.* **119**, 105794 (2023).

[34] Abdollahzadeh, B., Soleimanian Gharehchopogh, F. & Mirjalili, S. Artificial gorilla troops optimizer: A new nature-inspired metaheuristic algorithm for global optimization

[35] Chauhan, S., Vashishtha, G. & Kumar, A. A symbiosis of arithmetic optimizer with slime mould algorithm for improving global optimization and conventional design problem. *J. Supercomput.* (2021). doi:10.1007/s11227-021-04105-8

[36] Zhang, Y., Du, S. & Zhang, Q. Improved Slime Mold Algorithm with Dynamic Quantum Rotation Gate and Opposition-Based Learning for Global Optimization and Engineering Design Problems. *Algorithms* **15**, (2022).

[37] Zhu, Y. *et al.* Intelligent fault diagnosis of hydraulic piston pump combining improved LeNet-5 and PSO hyperparameter optimization. *Appl. Acoust.* **183**, 108336 (2021).

[38] Zhu, Y. *et al.* Acoustic signal-based fault detection of hydraulic piston pump using a particle swarm optimization enhancement CNN. *Appl. Acoust.* **192**, 108718 (2022).

[39] Tang, S., Zhu, Y. & Yuan, S. Intelligent fault diagnosis of hydraulic piston pump based on deep learning and Bayesian optimization. *ISA Trans.* **129**, 555–563 (2022).

problems. *Int. J. Intell. Syst.* **36**, 5887–5958 (2021).